\documentclass[10pt,a4paper,twocolumn]{article}
\usepackage{f1000_styles}
\usepackage{graphicx,wrapfig}
\usepackage{enumitem}
\usepackage{hyperref}
\usepackage{multirow}
\usepackage{multicol}
\usepackage{ulem}
\hypersetup{colorlinks,linktocpage,linkcolor={darkblue}, citecolor={darkblue}, urlcolor={darkblue}}

\usepackage{graphics,rotating}
\usepackage{xcolor}
\usepackage{multirow}
\usepackage{longtable}
\usepackage{booktabs}
\usepackage{subfig}
\usepackage{float,capt-of}
\usepackage[english]{babel}

\usepackage{atlast}

\definecolor{darkblue}{HTML}{45596E}

\newcommand{\ci}{[C\,{\footnotesize I}]}

\newcommand{\hi}{H\,{\footnotesize I}}
\newcommand{\cione}{[C\,{\footnotesize I}]$(^3P_1\,-\,^{3}P_0)$}
\newcommand{\citwo}{[C\,{\footnotesize I}]$(^3P_2\,-\, ^{3}P_1)$}

\newcommand{\cii}{[C\,{\footnotesize II}]\,}
\newcommand{\oiii}{[O\,{\footnotesize III}]\,}
\newcommand{\nii}{[N\,{\footnotesize II}]\,}
\newcommand{\niii}{[N\,{\footnotesize III}]\,}

\newcommand{\HII}{H$_2$}

\newcommand{\My}{${\rm M_\odot\, yr^{-1}} $}

\begin{document}
\pagestyle{fancy}

\title{Atacama Large Aperture Submillimeter Telescope (AtLAST) Science: The hidden circumgalactic medium}
\author[1,2]{Minju M. Lee}
\author[3]{Alice Schimek}
\author[3]{Claudia Cicone}
\author[4]{Paola Andreani}
\author[4]{Gerg\"{o} Popping}
\author[5]{Laura Sommovigo}
\author[6]{Philip N. Appleton}
\author[7,8]{Manuela Bischetti}
\author[9]{Sebastiano Cantalupo}
\author[10]{Chian-Chou Chen}
\author[11,12]{Helmut Dannerbauer}
\author[4]{Carlos De Breuck}
\author[13,7,8,14]{Luca Di Mascolo}
\author[15]{Bjorn H.C. Emonts}
\author[4,11,12]{Evanthia Hatziminaoglou}
\author[9]{Antonio Pensabene}
\author[1,16]{Francesca Rizzo}
\author[17,18,19]{Matus Rybak}
\author[3]{Sijing Shen}
\author[20]{Andreas Lundgren}
\author[21]{Mark Booth}
\author[21]{Pamela Klaassen}
\author[4]{Tony Mroczkowski}
\author[22]{Martin A. Cordiner}
\author[23,24]{Doug Johnstone}
\author[4]{Eelco van Kampen}
\author[25,26]{Daizhong Liu}
\author[27]{Thomas Maccarone}
\author[28,29]{Am\'{e}lie Saintonge}
\author[30]{Matthew Smith}
\author[31]{Alexander E. Thelen}
\author[3,32]{Sven Wedemeyer}

\affil[1]{Cosmic Dawn Center (DAWN), Denmark}
\affil[2]{DTU-Space, Technical University of Denmark, Elektrovej 327, DK2800 Kgs. Lyngby, Denmark}
\affil[3]{Institute of Theoretical Astrophysics, University of Oslo, PO Box 1029, Blindern 0315, Oslo, Norway}
\affil[4]{European Southern Observatory, Karl-Schwarzschild-Stra\ss e 2, 85748 Garching bei M\"{u}nchen, Germany}
\affil[5]{Center for Computational Astrophysics, Flatiron Institute, 162 5th Avenue, New York, NY 10010, USA}
\affil[6]{Caltech/IPAC, MC 314-6, 1200 E. California Blvd., Pasadena, CA 91125, USA}
\affil[7]{Astronomy Unit, Department of Physics, University of Trieste, via Tiepolo 11, Trieste 34131, Italy} 
\affil[8]{INAF -- Osservatorio Astronomico di Trieste, via Tiepolo 11, Trieste 34131, Italy}
\affil[9]{Department of Physics, University of Milan Bicocca, Piazza della Scienza 3, I-20126 Milano, Italy}
\affil[10]{Academia Sinica Institute of Astronomy and Astrophysics (ASIAA), No. 1, Sec. 4, Roosevelt Road, Taipei 10617, Taiwan}
\affil[11]{Instituto de Astrof\'{i}sica de Canarias (IAC), E-38205 La Laguna, Tenerife, Spain}
\affil[12]{Universidad de La Laguna, Dpto. Astrof\'{i}sica, E-38206 La Laguna, Tenerife, Spain}
\affil[13]{Laboratoire Lagrange, Université Côte d'Azur, Observatoire de la Côte d'Azur, CNRS, Blvd de l'Observatoire, CS 34229, 06304 Nice cedex 4, France}
\affil[14]{IFPU -- Institute for Fundamental Physics of the Universe, Via Beirut 2, 34014 Trieste, Italy}
\affil[15]{National Radio Astronomy Observatory, 520 Edgemont Road, Charlottesville, VA 22903, USA}
\affil[16]{Niels Bohr Institute, University of Copenhagen, Jagtvej 128, 2200 Copenhagen N, Denmark}
\affil[17]{Leiden Observatory, Leiden University, PO Box 9513, 2300-RA Leiden, The Netherlands}
\affil[18]{Faculty of Electrical Engineering, Mathematics and Computer Science, Delft University of Technology, Mekelweg 4, 2628 CD Delft, The Netherlands}
\affil[19]{SRON - Netherlands Institute for Space Research, Niels Bohrweg 4, 2333 CA Leiden, The Netherlands}
\affil[20]{Aix Marseille Univ, CNRS, CNES, LAM, Marseille, France}
\affil[21]{UK Astronomy Technology Centre, Royal Observatory Edinburgh, Blackford Hill, Edinburgh EH9 3HJ, UK}
\affil[22]{Astrochemistry Laboratory, Code 691, NASA Goddard Space Flight Center, Greenbelt, MD 20771, USA.}
\affil[23]{NRC Herzberg Astronomy and Astrophysics, 5071 West Saanich Rd, Victoria, BC, V9E 2E7, Canada}
\affil[24]{Department of Physics and Astronomy, University of Victoria, Victoria, BC, V8P 5C2, Canada}
\affil[25]{Max-Planck-Institut f\"{u}r extraterrestrische Physik, Giessenbachstrasse 1 Garching, Bayern, D-85748, Germany}
\affil[26]{Purple Mountain Observatory, Chinese Academy of Sciences, 10 Yuanhua Road, Nanjing 210023, China}
\affil[27]{Department of Physics \& Astronomy, Texas Tech University, Box 41051, Lubbock TX, 79409-1051, USA }
\affil[28]{Department of Physics and Astronomy, University College London, Gower Street, London WC1E 6BT, UK}
\affil[29]{Max-Planck-Institut f\"ur Radioastronomie (MPIfR), Auf dem H\"ugel 69, D-53121 Bonn, Germany}
\affil[30]{School of Physics \& Astronomy, Cardiff University, The Parade, Cardiff CF24 3AA, UK}
\affil[31]{Division of Geological and Planetary Sciences, California Institute of Technology, Pasadena, CA 91125, USA.}
\affil[32]{Rosseland Centre for Solar Physics,  University of Oslo, Postboks 1029 Blindern, N-0315 Oslo, Norway}

\maketitle
\thispagestyle{fancy}

\clearpage
\begin{abstract}

Our knowledge of galaxy formation and evolution has incredibly progressed through multi-wavelength observational constraints of the interstellar medium (ISM) of galaxies at all cosmic epochs. However, little is known about the physical properties of the more diffuse and lower surface brightness reservoir of gas and dust that extends beyond ISM scales and fills dark matter haloes of galaxies up to their virial radii, the circumgalactic medium (CGM). New theoretical studies increasingly stress the relevance of the latter for understanding the feedback and feeding mechanisms that shape galaxies across cosmic times, whose cumulative effects leave clear imprints into the CGM. Recent studies are showing that a -- so far unconstrained -- fraction of the CGM mass may reside in the cold ($T<10^4$~K) molecular and atomic phase, especially in high-redshift dense environments. These gas phases, together with the warmer ionised phase, can be studied in galaxies from $z\sim0$ to $z\sim10$ through bright far-infrared and sub-millimeter emission lines such as \cii~158$\mu$m, \oiii~88~$\mu$m, \ci~609$\mu$m, \ci~370$\mu$m, and the rotational transitions of CO. Imaging such hidden cold CGM can lead to a breakthrough in galaxy evolution studies but requires a new facility with the specifications of the proposed Atacama Large Aperture Submillimeter Telescope (AtLAST). In this paper, we use theoretical and empirical arguments to motivate future ambitious CGM observations with AtLAST and describe the technical requirements needed for the telescope and its instrumentation to perform such science.
\end{abstract}

\clearpage

\section*{\color{OREblue}Keywords}
Galaxies: circumgalactic medium; Galaxies: intergalactic medium; Galaxies: ISM; Galaxies: evolution; Submillimeter: galaxies; Radio lines: galaxies.

\pagestyle{fancy}

\section*{Plain language summary} 
The paper aims to demonstrate the need for a new large aperture (50 m), single-antenna telescope receiving submillimeter and millimeter (hereafter sub-mm)\footnote{We refer to the whole submillimeter and millimeter range as sub-mm, implying the wavelength between 0.35 mm and 10 mm.} wavelength light from a high elevation site in the Atacama desert in Chile, named the Atacama Large Aperture Submillimeter Telescope (AtLAST). Here, we particularly focus on the science case of the so-called circumgalactic medium (CGM). This gaseous component exists beyond the scale of the matter that lies between stars in a galaxy (the interstellar medium, ISM) but still within the gravitationally bounded region of a galaxy. Our understanding of galaxies has so far been based on observations that focus on the ISM, but theory shows that observing the CGM may help us solve crucial open questions in the field of galaxy formation and evolution. Indeed, the properties of the CGM carry the vital imprints of the physical mechanisms that shape galaxies, specifically the powerful winds driven by newly formed stars and by supermassive black holes, and the incoming gas flows from the large-scale structure of the Universe that provide galaxies with their fuel to form stars. Despite its crucial role, little is known about the CGM, particularly its cold and dense gas content, because none of the current sub-mm telescopes enables such observations. We illustrate that exploring the hidden cold CGM components is an urgent task in the coming decades and evaluate how feasible this science case is, based on our current knowledge. We suggest a set of telescope parameters and instrumentation for AtLAST to achieve such key science goals of probing the cold CGM. Finally, we discuss expected synergies with current and future telescopes.

\section{Introduction: the breakthrough potential of AtLAST for CGM observations}\label{sec:intro}

~~The gaseous reservoir surrounding galaxies, enriching their dark matter halos beyond the scales of the interstellar medium (ISM) and up to their virial radii, is loosely defined\footnote{We do not impose a strict definition of the CGM. In our loose definition, the CGM includes the matter at the boundary between ISM and CGM. For example, we call CGM gas components a few kpc from the disc of at $z=6$ star-forming galaxy with stellar mass of $\log{M_{\star}/M_{\odot}}\sim 10$, or at $\sim10$ kpc scales in a $z=2$ system with $\log{M_{\star}/M_{\odot}}\sim 11$, which 
are yet beyond the extent of rest-frame optical sizes (e.g., \citealt{vanderwel2014,Ormerod2024}).} as the circumgalactic medium (CGM). 
The CGM is a critical component of the galaxies' ecosystem because it interfaces their disc reservoirs with their external supply of gas from the intergalactic medium (IGM). The processes that occur on CGM scales, such as galactic outflows, cosmological inflows and galactic fountains, galaxy mergers, tidal interactions, ram pressure stripping, cooling, heating, and photoionisation from active galactic nuclei (AGN), can affect the galaxy's ability to sustain star formation.
The physical properties of the CGM, e.g. gas and dust content, metal content, temperature, size, density distribution, and morphology, are the fossil imprints of past feedback mechanisms, galaxy interactions, cosmic accretion, and so are crucially linked with the galaxy evolution. 
For example, the CGM offers a unique opportunity to understand how galactic outflows (hereafter, outflows refer to galactic outflows) operate in galaxies. Indeed, while the stellar masses of galaxies are largely agnostic (at low redshift) to whether galaxy growth is regulated by violent, intermittent episodes of feedback or by a gentler and more continuous process, the physical conditions of the CGM are expected to differ strongly in different feedback scenarios \citep{CrainvandeVoort,2023CGMTheory}.

~~Observationally, the nature of the CGM is highly elusive. 
This is due to two of its intrinsic characteristics: (i) the physical scales of interest for CGM studies range from hundreds of kpc (the virial radius of a Milky Way mass galaxy is $\sim$200~kpc; e.g., \citealt{Grcevich2009, Fang2013}) down to sub-kpc scales, which is the expected size of the clumps that enrich this medium; 
(ii) the considerable dynamical range needed to simultaneously capture the diffuse, large-scale CGM reservoir and the compact, high surface brightness ISM component implies that observations targeting the latter will hardly be suitable to study the former. 
In addition to these two critical observational limitations, a highly multi-phase nature of the CGM is emerging from recent observations (see Section~\ref{sec:obs_constr}), showing that significant amounts of cold molecular gas at $T\lesssim100$~K can exist on CGM scales in addition to the $T\sim10^{5-6}$~K warm/hot gas phase components that are believed to be dominant (see e.g. \citealt{Tumlinson2017, Nicastro2023} and companion AtLAST case study by Di Mascolo et al. 2024 (in preparation). 

~~Despite being highly elusive from current observing facilities, a handful of robust results suggest the existence of a cold CGM component, which we summarised in Section~\ref{sec:obs_constr}. 
Unfortunately, current facilities do not allow us to understand whether the massive molecular CGM reservoirs discovered at $z\sim2$ and extended \cii emissions at $z>4$ (see Section~\ref{sec:obs_constr}) are due to unresolved satellites and merging companions, or rather they trace clump condensation out of the CGM and/or genuinely diffuse and large-scale molecular gas streams in inflow or outflow. 
To understand this, we need to study cold CGM with better sensitivity by tracing multiple lines that allow us to constrain the physical properties of the CGM and pin down the possibly missed extended diffuse emission. 

\begin{figure*}[htb]
    \centering
    \includegraphics[width=0.93\textwidth]{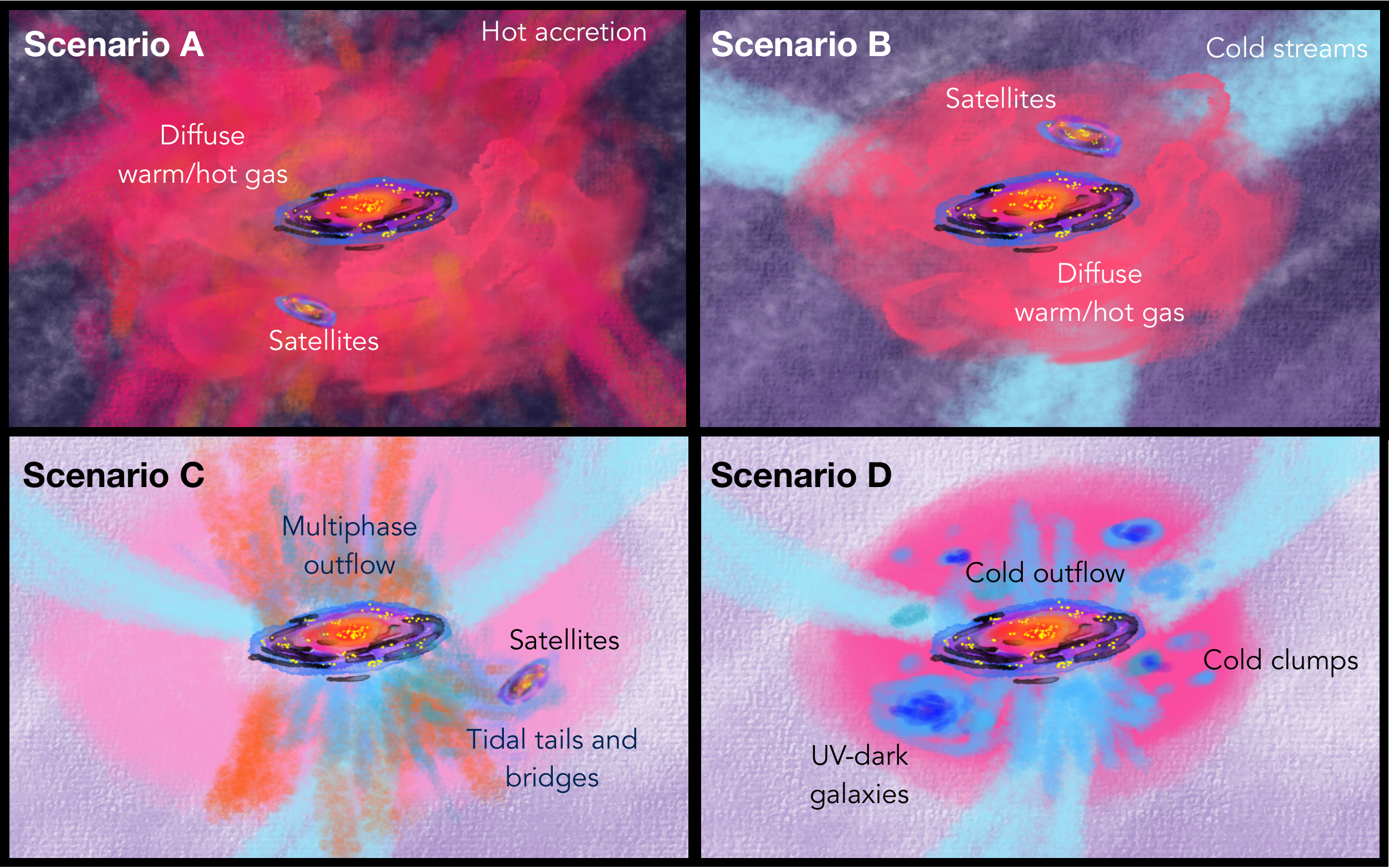}
    \caption{Cartoon illustrating different scenarios for the CGM composition, inspired by Figure~1 in \cite{Tumlinson2017}. The galaxy is illustrated as a star-forming (yellow dots) disc partially shrouded by dust (black) and a central bulge (red) surrounded by loosely defined CGM components as labelled in each scenario.
    \textbf{Scenario A}: A galaxy and its satellite galaxies are surrounded by diffuse hot/warm CGM and IGM components fed by hot mode accretion in varying tones of red colour, with no diffuse cold and dense gas contribution. 
    In this scenario, any detection of cold gas on CGM scales is expected to be due entirely to the ISM of satellites. 
    \textbf{Scenario B}: The accretion from the IGM is in cold mode (sky blue). The streams cannot fully penetrate the dark matter halo so most of the CGM is hot/warm phase gas except for the ISM contribution from satellites. 
    \textbf{Scenario C}: In Scenario C, a galaxy is fed by cold gas accretion along the filaments and cold accretion reaches the central galaxy contributing to the CGM. Moreover, the central galaxy has launched a powerful multi-phase outflow extending by several kpc that has both cold (blue) and warm/hot (orange) gas. An additional contribution to the cold CGM is given by tidal tails formed from galaxy interactions. 
    \textbf{Scenario D}: Cold accreting filaments (sky blue), cold galactic outflows and fountains (blue), optically dark satellites (large blue clump), and cold condensation out of the CGM itself (cold CGM clumps; smaller blue clump) lead to a significant fraction of the CGM to be in a cold and relatively dense phase. 
    Current observational and theoretical work suggests that scenarios C and D may be dominant at high-$z$. The CGM of local galaxies may resemble more that of Scenarios A and B, although evidence for massive cold molecular outflows in local ultraluminous infrared galaxies and cold tidal material between galaxies in groups suggests that some elements of Scenarios C and D may also be found in the $z\sim0$ Universe.
    }
    \label{fig:keyquestions}
\end{figure*}

~~The existence of a (possibly very massive) neutral atomic and molecular CGM component implies that none of the future purpose-built optical/UV band instruments (such as BlueMUSE and CUBES on the ESO Very Large Telescope, see e.g. \citealt{BlueMUSE_sciencecase, Evans+18})
will ever be able to solve the CGM question without synergetic sub-mm observations.
A sensitive, large field of view (1-2 deg), large aperture (50-m) single dish facility, housing up to six state-of-the-art multi-beam instruments, enabling continuum and spectral line imaging from arcsec to degrees scales such as the Atacama Large Aperture Submillimeter Telescope 
(AtLAST, \citealt{Mroczkowski2024, Mroczkowski2023, Ramasawmy2022, Klaassen+20})
will open up exploration of the uncharted territory of the cold CGM phase. 
This case study seeks to motivate the AtLAST endeavour with the scope of pursuing ambitious observations of the hidden cold CGM of galaxies. 
Imaging the hidden cold CGM with AtLAST is a high-reward science case that can lead to a breakthrough in galaxy evolution studies. 

~~In Figure~\ref{fig:keyquestions} we show a cartoon illustrating various scenarios for the composition of the CGM, drawn based on our current understanding.
We lay out four scenarios (A, B, C and D that may not be necessarily mutually exclusive) for possible states of the CGM.
Each of these scenarios is valid and possibly relevant for a different galaxy evolutionary stage: determining such a link is the main goal of CGM observations. 
From the point of view of sub-mm observations, any proof of potential scenarios described in Figure~\ref{fig:keyquestions} will open a new avenue to studies of galaxy formation and evolution.
For example, even in the most conservative case of a cold CGM component that is made entirely of the unresolved ISM of optically faint (or optically dark) galaxy companions and dominated by only hot accretion mode (Scenario A), sub-mm observations hold promise for directly imaging such hidden components confined in the ISM of individual galaxies, whose existence was previously unaccounted for. 
Cold mode accretion (e.g. \citealt{Dekel+09, Keres+09}, shown as sky blue in Scenario B, C, D in Figure~\ref{fig:keyquestions}), massive molecular outflows on galactic scales of several kpc (Scenario D; e.g., \citealt{Veilleux+20}), tidal interactions and satellites (Scenario C; e.g., \citealt{TC2021}) are additional mechanisms that can populate galaxy haloes with cold and dense gas.
In particular, the nature of accreting filaments is still unconstrained: the streams can be gravitationally unstable and thus fragment \citep{Mandelker+18}, forming denser (possibly star-forming) clumps that become part of the CGM reservoir (Scenario D). 
As a note, constraining the physical drivers that populate such components depicted in Figure~\ref{fig:keyquestions} will eventually require multiple probes across all electromagnetic waves and data combination with sub-mm interferometric observations (see Section~\ref{sec:synergy}). 
However, pursuing a large-scale search of cold CGM components in galaxies at different cosmic epochs is the next big challenge of galaxy formation and evolution studies and an urgent task to be accomplished.
It is intrinsically tied to the development of AtLAST, which is the only facility on the horizon that enables cold CGM science. 
AtLAST's capability will allow us to map the extended, hidden cold CGM with exquisite sensitivity, which current sub-mm facilities largely miss.

~~In the remainder of this section, we summarise the most crucial pending questions that could be addressed through the proposed CGM observations with AtLAST. 
In Section~\ref{sec:obs_constr} we summarise the results from a few pilot observational studies of the cold CGM; in Section~\ref{sec:simulations} we discuss the challenges faced by computational studies attempting to provide theoretical predictions on the cold CGM. In Section~\ref{sec:atlast} we illustrate what could be done with AtLAST. In Section~\ref{sec:tech_req} we describe the technical requirements needed for AtLAST to pursue the proposed observations. We conclude this paper in Section~\ref{sec:synergy} with a brief discussion of the role of other current or planned facilities in complementing AtLAST for synergistic CGM observations.

\subsection{What can we learn from CGM observations?}
\vspace{1pt}
~~There are at least three main open questions in the field of galaxy formation and evolution that require dedicated CGM observations: i) The nature of cosmic accretion, ii) missing baryon and missing constituents, and iii) feedback mechanisms. We describe the details in the following.

\subsubsection*{I. The nature of cosmic accretion}
~~The last decade achieved a huge success in constraining the cold ISM in the distant universe (up to redshift $z\sim6$; e.g., \citealt[]{Scoville2013, Tacconi2018, Tacconi2020, Zanella2018, Walter2020, Vallini2024}).
The evolution of the cold gas content of galaxies and their molecular gas depletion time-scales (defined as, $\rm \tau_{dep}\equiv M_{mol}/SFR$, i.e. the time it takes a galaxy to consume its H$_2$ gas reservoir at the current star formation rate, SFR) explains the overall evolution of the cosmic star-formation rate density of the Universe, which peaks at $z=1-3$ (\citealt[]{Tacconi2020} for a review and references therein).
However, especially at $z>1$, the gas depletion time scale at a given star-formation rate (a few hundred Myr up to a couple of Gyr) is shorter than the Hubble time ($\gtrsim$5-8 Gyr) at $z>1$. 
This strongly requires a sustained gas accretion to keep up with the star-formation rates observed in galaxies -- otherwise, galaxies would quench too soon.
Therefore, gas accretion, either from cosmic streams in the IGM or from a reprocessed CGM phase, is needed to explain the prolonged star formation activity of galaxies across cosmic times.
Two unanswered questions are (i) how and in what phase such gas is accreted onto galaxies, and (ii) what physical mechanisms regulate galaxy feeding.

\subsubsection*{II. The missing baryon problem}
\begin{figure*}[tbp]
    \centering
    \includegraphics[width=0.93\textwidth]{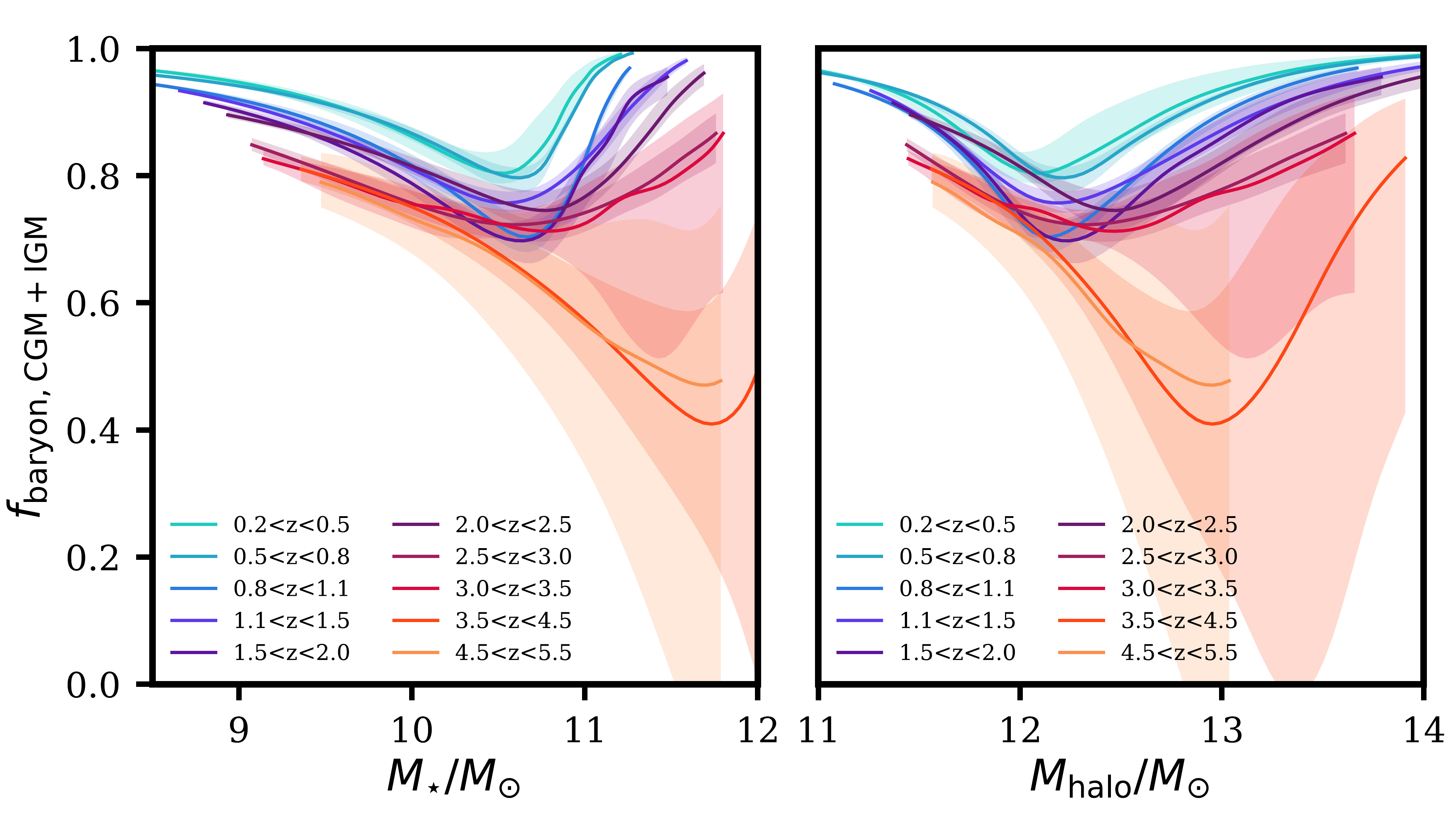}
    \caption{The inferred baryon fraction in the CGM and IGM of star-forming main sequence galaxies as a function of stellar mass and halo mass, based on the observational constraints of the following: (i) a baryon fraction of 16\% \citep{PlanckCollaboration2018}, (ii) the stellar-to-halo mass relation constructed from COSMOS2020 catalogue (\citealt{Shuntov2022}), and (iii) the cold molecular gas content (in the ISM) of main-sequence galaxies at given mass and redshift from \citet{Tacconi2018}. See the description of the calculation in the main text. The figure shows that a large fraction of baryons is expected to reside in the CGM.
    The curves are truncated in the low-mass and high-mass ends owing to the lack of observational constraints in both mass regimes from the stellar-to-halo mass relation. Uncertainties are shown as the shaded area, which is likely underestimated; this is because we extrapolated the gas content relation (iii) to match with the stellar (and thus halo) mass ranges of the relation (ii), but both low ($M_{\star}\lesssim10^{10}\, M_{\odot}$) and high mass ($M_{\star}\gtrsim10^{11}\, M_{\odot}$) ends are hardly constrained observationally.  We need better constraints of this baryon budget in the CGM from \textit{direct} observations using AtLAST.}
    \label{fig:fbaryon_cgm}
\end{figure*}

~~There is a long-standing discrepancy between the expected baryons from the cosmic mean baryon fraction, $f_{\rm b}\approx 0.16$ \citep{PlanckCollaboration2018} for a given halo, and the observed baryonic mass components within galaxies (stars and ISM), including the Milky Way (e.g. \citealt{Maller2004, Anderson2010, Crain2010, McGaugh2010, Feldmann2013, Tumlinson2013, Tumlinson2017, Schaller2015, vandeVoort2016, Suresh2017, Bregman2018, Nicastro2018}); we observe less mass than we expect. 
This missing mass is comparable to or even more significant than what is observed within galactic discs.
Figure~\ref{fig:fbaryon_cgm} shows the expected fraction of baryons in the CGM (and IGM) for a star-forming main-sequence galaxy, inferred from the latest observational constraints of the following: cosmic baryon fraction ($f_{\rm b} =$ 16\%; \citealt{PlanckCollaboration2018}), the stellar-to-halo mass ratio as a function of redshift \citep{Shuntov2022} and the ISM content\footnote{Mainly the molecular gas, here we did not take into account the contribution of the atomic gas (\hi)} at given stellar mass and redshift \citep{Tacconi2018}. 

~~To calculate the fraction, we take a few steps of assumption in the following. For a given halo mass ($M_{\rm halo}$) and redshift (z), we estimate the stellar mass ($M_{\star}$) based on the redshift-dependent stellar-to-halo ratio, and the ISM content ($M_{\rm ISM}$) from the gas fraction ($\mu = M_{\rm ISM}/M_{\star}$) using the gas scaling relation, which is a function of redshift, deviation from the main-sequence, stellar mass (and weakly on galaxy size), assuming a galaxy being a star-forming main sequence galaxy. We then subtract the contribution of the star and ISM from the expected baryon mass ($M_{\rm bar} = M_{\rm halo} \times f_{\rm b}$) to get the remaining baryon mass not included in stars and ISM (i.e, in the CGM and IGM).

~~Although such inference of the CGM and IGM fractions include many assumptions with associated large uncertainties, it indeed suggests that $\gtrsim50\%$ of "missing" baryons are located in the outer regions (e.g., $r\gtrsim$10 kpc at $z\sim 1$ for $\gtrsim 10^{11}\, M_{\odot}$ galaxies; $r\gtrsim$ a few kpc at $z\sim5$ for $M_{\star}\gtrsim 10^{10}\, M_{\odot}$, where $r$ is the radius), namely in the CGM, at all redshift and all stellar mass ranges (see also reviews by \citealt{Tumlinson2017, Tacconi2018, Peroux2020}).
The contribution of cold gas to the total mass budget of CGM is far from being constrained, and gas in such a phase could be readily available for star formation in the galaxy.

\subsubsection*{III. The impact of feedback from star formation and AGN on galaxy evolution}

~~Feedback is one of the most critical components of our galaxy evolution theory, without which semi-analytic models and simulations cannot reproduce the most basic observables of galaxies, such as stellar mass function at $z\sim0$ (e.g., \citealt{White1991, Springel2003}, see also recent review by \citealt{CrainvandeVoort}).
The observed stellar-to-halo mass relation \citep{White1978, White1991, Balogh2001}, shows that low and high-mass galaxies are less efficient in turning baryons into stars, and the peak efficiency is at $M_{\rm halo}\sim10^{12}\, M_{\odot}$.
The inefficient transformation of baryons into stars in low and high stellar (or halo) mass regimes are successfully explained by the feedback mechanisms that are dominant in different mass regimes: feedback from star formation in low-mass galaxies and feedback from accreting supermassive black holes (SMBH) in massive galaxies and galaxy clusters (e.g., \citealt{Dekel1986, Silk1998, Efstathiou2000, Springel2003, Springel2005d, DiMatteo2005}). 
In simulations, these feedback agents could act on galaxies in the form of powerful galactic-scale outflows, halo heating, suppression of accretion, injection of turbulence into the ISM, etc.

~~However, the exact forms and consequences of feedback from star formation, and active galaxy nuclei (AGN), are poorly constrained observationally \citep{Harrison+18, Cicone+18NatAS}. 
The lack of solid and robust observational constraints implies that most models and simulations must fine-tune their feedback parameters, mostly by calibrating their stellar-to-halo mass ratio to $z=0$ observations.
One of the observational challenges is that some feedback mechanisms, such as halo heating and the following suppression of accretion \citep{Fabian12}, could act on timescales ($\gg$ Gyr) much longer than those over which the processes driving the feedback remain detectable.
For example, AGN outflows are detectable for much longer than the AGN flickering timescale (e.g., \citealt{Herrera-Camus+20, Bischetti2019}).
Further, there is a lack of observational evidence (with only a few exceptions; see \citealt{Circosta+21, Cresci+23}) that the feedback is instantly effective at ISM scales of a few kpc.

~~CGM observations will allow us to understand the cumulative effect of feedback mechanisms and the dominant mechanisms that shape the host galaxy and its evolution.
Galactic outflows and fountains transport gas and metals into the CGM \citep{Tumlinson2011, Rupke+19, Travascio+20, Vayner+23} and beyond, mixing it with existing CGM gas and also possibly polluting the IGM \citep{Oppenheimer+Dave06,Mitchell+Schaye22}.
Therefore, the properties of the CGM, for instance, the metallicity, radial distribution, and breakdown of multiphase components, are shaped differently by different physical mechanisms.
Further, the records of past feedback activities could be traced by following the outflows' relics on much longer spatial scales (and so, time scales) than enabled by ISM observations.
A notable example is that of the ``Fermi (or eRosita) Bubbles'' in the Milky Way's CGM, which are a clear signature of past feedback activity in our Galaxy that has otherwise left no clear imprints on ISM scales \citep{Su+10, Predehl+20, Yang2022}. 
Molecular gas clouds in the outflow are also detected by \cite{DiTeodoro+20} in CO(2-1) using APEX in conjunction with the Southern Fermi Bubble. The latter has an angular size of $\rm 50~deg\times 15~deg$, and such scale can be probed in the sub-mm only by a facility such as AtLAST.\\

~~Investigating the physics driving the processes shaping the CGM of galaxies requires sampling the whole CGM, from its hot ($\gtrsim10^5$ K), warm ($\sim10^4$ K) to cold phases ($\lesssim10^3$ K) (drawn respectively in red, orange and blue colours in Figure~\ref{fig:keyquestions}). 
This effort enforces the synergy of multi-wavelength observations highlighted in Section~\ref{sec:synergy}.

\vspace{1pt}

\section{Observational constraints from nearby to distant universe}\label{sec:obs_constr}
~~In this section, we briefly review the current observational constraints of the cold CGM.
The existence of a cold and dense CGM component has never been predicted theoretically, mostly due to challenges in modelling, which we describe in Section~\ref{sec:simulations}.
However, in recent years, we have gathered pilot observational constraints pointing towards a significant amount of cold gas, even in the molecular phase, extending beyond the expected scales of galaxy discs, in some extreme cases up to hundreds of kpc around massive galaxies at $z\sim2$. Before discussing these extraordinary results at high redshift (which probably reflects a very different Universe with a richer environment than $z\sim0$), we shortly summarise current observational evidence for extended cold gas reservoirs in local galaxies, suggesting that we are missing components in the outer ISM and, perhaps, CGM. 
As a note, the review does not aim to provide a thorough list of the literature of relevant studies but aims to address the need for AtLAST to probe the cold CGM.

\begin{figure*}
    \centering
    \includegraphics[width=\textwidth]{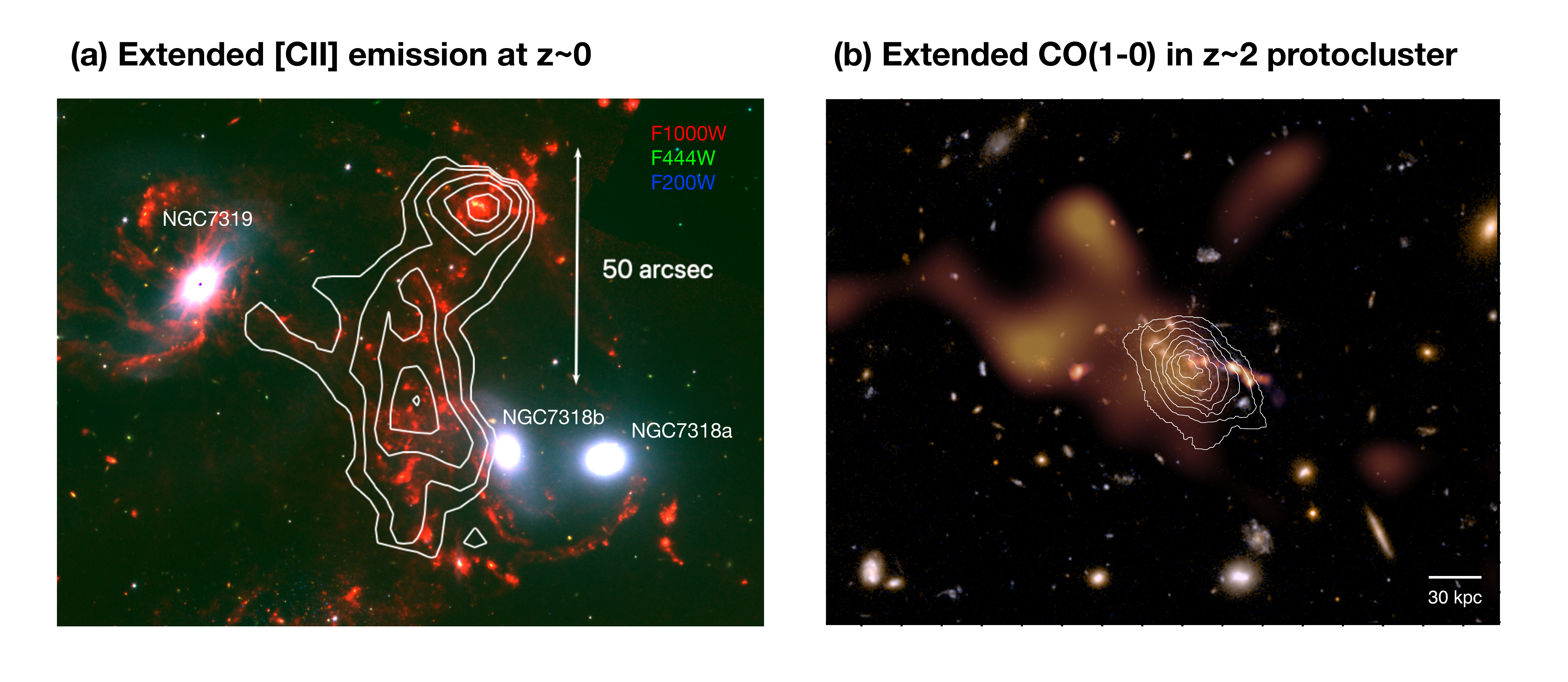}
    \caption{Examples of sources showing extended ($\gtrsim$ 10s arcsec) cold gas emission in the CGM at different redshifts. (a) \cii\, line emission obtained by the {\it Herschel} Space Telescope in Stephan's Quintet at z = 0.02 (\citealt{Appleton+13, Appleton+17}, white contours) overlaid on the JWST Early Release Observations false-colour maps (F1000W/F444W/F200W; \citealt{Appleton+23}), and (b) Extended CO(1-0) emission (white contours, \citealt{Emonts2016}) in the $z\sim2$ Spiderweb protocluster overlaid on the (extended) SZ signal (orange, \citealt{DiMascolo2023}) and false-colour \textit{HST} image.}
    \label{fig:currentobs}
\end{figure*}

\subsection{The local  Universe: missed extended emission in the outer ISM and possibly CGM of $z\sim0$ galaxies}\label{sec:nearby}

~~At low- to intermediate-redshift ($z<1$), evidence of extended (beyond the central few kiloparsecs), molecular and atomic gas reservoirs are found around starburst galaxies \citep{Zschaechner+18, Contursi+13, Leroy+15, Levy+23}, AGN \citep{Maccagni+21, Ianjamasimanana+22}, normal disc galaxies (e.g., \citealt{Das2020}), and most frequently in ultra-luminous infrared galaxies (ULIRGs), which host both starbursts and AGN \citep{Cicone+14, Cicone+18, MontoyaArroyave+23, MontoyaArroyave2024}. The origin of the extended gas reservoir in the CGM could be past galaxy mergers and interactions and/or current or past star formation and AGN feedback episodes that launch multi-phase outflows that entrain atomic and molecular gas (see \cite{Veilleux+20} for a review on this topic).

~~Galaxy interactions could produce tidal tails and tidal tail dwarf galaxies (TDGs, see \citealt{Ploeckinger+18} for theoretical background, and \citealt{Duc+14, Kaviraj+12} for observational works). 
Gas tails formed by galaxy interactions, tidal forces and/or ram pressure stripping can reach extents of several hundreds of kpc, such as the  500~kpc long HI filament detected by \cite{Oosterloo+18} in the IC~1459 galaxy group, or the $\sim200$~kpc Magellanic Stream (hereafter, `MS'\footnote{A resolution was recently proposed to the International Astronomical Union by the IAU Inter-Commission C1-C3-C4 Working group on Ethnoastronomy and Intangible Astronomical Heritage to rename the main Milky Way satellites and their associated structures.  While waiting for a final decision on this, we will use the acronym `MS' throughout this paper.}), created by the interaction of the Milky Way with its satellites and representing the most massive HI CGM reservoir of our Galaxy (e.g., \citealt{Braun2004, D'Onghia+16}). 
Another spectacular example in the local volume is a 3~deg-wide VLA mosaic image of the M~81 triplet (including M~82, hosting a prototypical starburst-driven outflow) obtained by \cite{deBlok+18}, showing a $\sim100$~kpc size HI reservoir connecting all members of the galaxy group.
Extended CO~(1-0) emission is also discovered and coincides with the HI emission \citep{Krieger2021}.
Tidal tails and streams from interactions probably constitute most of the HI emission seen between the M~81 triplet members, in addition to the matter expelled by the M~82 outflow \citep{deBlok+18}. 
While HI~21cm imaging on scales of degrees can be easily achieved by SKAO pathfinders such as MeerKAT, there is no existing sub-mm facility that has sufficient surface-brightness sensitivity and field of view (FoV) to enable adequate follow-ups of such structures in CO, dust emission and other sub-mm tracers: only AtLAST could do it.

~~An interesting and famous case study for AtLAST to probe the impact of outflows with a large mapping area would be the galaxy NGC~253, where extended, diffuse, molecular components not associated with the galaxy disc are detected by several authors \citep{Walter+17, Krieger+19}. 
According to \cite{Bailin+11}, the CGM of NGC 253 is comparable in mass to that of the Milky Way and M31. 
In this case, the molecular outflow rate is a factor of $9-19$ higher than the SFR \citep{Walter+17}, hence exceeding previously proposed theoretical limits for standard starburst-driven winds (which never exceed a mass-loading factor, i.e. the ratio between outflow rate and SFR, of $\sim5$).
\citet{Zschaechner+18} also find evidence for new, diffuse, extra-planar CO components.
These features extend up to the edge of the mapping area ($40''$), partially overlap with known outflow streams, and appear to form a biconical structure. 
The current limitations to further investigation are the small field of view and low sensitivity on large angular scales. 
A sampling at high S/N of the whole galactic ecosystem of NGC~253 ($\gtrsim 8$ kpc, or $\gtrsim 8'$, given the extent of diffuse X-ray and dust emissions; \citealt{Pietsch2000, Strickland2002, Bauer2008, Kaneda2009}) would become possible with an instrument that fills the FoV of AtLAST. Interestingly, the HI gas reservoir of NGC~253 imaged by \cite{Lucero+15} with KAT-7 exhibits halo emission out to $10-14$~kpc (10-14 arcmin) perpendicular to the disc, with kinematic properties that are compatible with a galactic fountain origin. 

~~Going beyond the local volume, the luminous infrared galaxy, and quasar host NGC~6240, at $z=0.02448$ ($1''\sim0.5$~kpc) displays one of the best studied, most massive (M$_{mol}^{out}\sim 1.2\times10^{10}~M_{\odot}$), and most extended ($r>10$~kpc) molecular outflows \citep{Cicone+18}.
The galaxy is in an advanced merger stage and it hosts a pair of AGN with a $\sim1$~kpc separation.
The total CGM of NGC~6240, as revealed in H$\alpha$ by \cite{Yoshida+16} extends by at least 90~kpc in diameter (3~arcmin). 
Radio continuum images obtained with the Very Large Array (VLA) exhibit extended features that are not associated with optical emission and are interpreted as the relic of a past feedback event \citep{Colbert+94}. MeerKAT observations unveil a 100~kpc-size radio relic with a morphology reminiscent of a shock front, located at a distance of 200~kpc from the nucleus of NGC~6240 (Priv. Communication). 
Although ALMA observations of NGC~6240 pin down the emission faint as $\sim$ sub-mJy/beam in the outflow with a beam of $\sim1.2''$, given the extent of the near-IR continuum, H$\alpha$ emission, and radio continuum in this source, we are very likely missing key components of the diffuse emission inaccessible to current sub-mm facilities (Section~\ref{sec:atlast}).
We need a sensitive instrument that can map the low surface brightness over the large area of the sky at high S/N to reveal additional outflow and tidal tail components extending by several tens of kpc.

\subsection{Gas-rich brightest cluster galaxies and interacting galaxies: different gas conditions?}\label{sec:multiphase}

~~There is a growing body of evidence that large reservoirs of molecular gas ($\gtrsim$ $10^{10}\,M_{\odot}$) exist in some intermediate- to high-redshift brightest cluster galaxies (BCGs), extending up to tens of kpc, feeding the central star-formation (e.g., \citealt{McDonald+14, Russell+17, Fogarty+19, Castignani+20b, Castignani+22,  Dunne+21}).
These results are the opposite cases of what is pictured in the core of dense clusters in the local universe, where gas-poor, quenched massive galaxies are typically found (e.g., \citealt{Haynes1984, Giovanelli1985, Denes2014, Vollmer2008, Boselli2014, Zabel2019}).
Such systems are unique laboratories to study the cooling flow-regulated star-formation activity at different stages of galaxy clusters.

~~Further, current observations of the CGM in several galaxy clusters guide us on additional mechanisms that we should take into account to understand galaxy evolution, requiring multi-phase probes.
A very instructive example is MACS1931-26 \citep{Fogarty+19} where in its CGM the gas-to-dust ratio ($\sim10-25$) is much lower than the Galactic value ($150$) and similar to the Phoenix cluster BCG \citep{Russell+17}. 
The unusual ratio suggests that a potentially large \HII\ gas reservoir is missed by CO observations, perhaps due to its dissociation by highly energetic particles (HEP) like Cosmic Rays (\citealt{Ferland+08, Ferland+09}), with possible dust and gas thermally decoupling. These conditions are different from what is seen in galaxy ISM.
Cosmic ray-irradiated CO-dark molecular gas can be revealed by using the atomic Carbon lines at rest-frame 492~GHz and 809~GHz \citep{Papadopoulos+18}, which, for the local universe, can be observed only from a high dry site such as the Chajnantor Plateau.
The CO gas excitation suggests that the temperature distribution of the gas differs across tens of kpc, indicating various energy sources at work in this environment (Ghodsi et al, 2024, submitted).
Hence, to constrain the physical condition of the CGM, studies of multiple emissions lines probing different gas phases (i.e., molecular, atomic, neutral, and ionised phases) and alternative tracers of H$_2$ gas are crucial, besides the initial cold CGM detection experiment. 

~~Another enlightening case is an interacting group of galaxies in the northern hemisphere, Stephan's quintet (Figure~\ref{fig:currentobs} (a)) at z=0.02, which shows extended emission from tidal streams in all phases (from X-ray to cm).
Multiple lines of the same atomic/molecular species are mapped, constraining the physical conditions of tidally disturbed emissions, and shock-heated regions in the CGM (e.g., \citealt{Gao+00, Appleton+06, Cluver+10, Guillard+12, Appleton+13, Guillard+22, Appleton+23, Xu+22, Cheng+23}).
Extensive studies with multiple emission lines demonstrate that molecular hydrogen (\HII) is formed along with dust in a region where hot X-ray plasma is detected in the CGM.
The dissipation of mechanical energy through shocks and turbulence, caused by the interaction of a high-speed intruder with group-wide gas, probably drives the powerful mid-IR H$_2$ and far-IR \cii\, emission. Cosmic ray heating in this group-wide multi-phase medium is unlikely (e.g., \citealt{Guillard+09, Appleton+13}). 
The discovery and future exploration of systems like Stephan's Quintet, containing pristine extended regions of turbulently-excited intergalactic gas, will lead to a better understanding of the fundamental physical mechanisms of galaxy evolution. This includes the importance of turbulence and gas-phase mixing in enhancing or inhibiting star formation as gas collapses to form stars in the early universe.

\subsection{Yet to be unveiled CGM emission in the distant Universe: AGN, star-forming galaxies, and proto-clusters}\label{sec:highz}
~~Studies of the distant universe ($z\gtrsim1$), around the peak of cosmic star-formation activity and beyond, provide lines of evidence that large cold reservoirs exist outside the galactic discs. 
Overdense environments and protoclusters show a promisingly molecular gas-rich CGM observed in CO \citep{Emonts2013, Dannerbauer2017, Ginolfi2017, Emonts2016, Emonts+18, Li2023, Chen2024}.
For example, a 100~kpc-long stream of cold molecular gas was detected at 5.7$\sigma$ to connect to the $z\sim4$ radio galaxy 4C 41.17 using ALMA observations of the \cione (hereafter, \ci(1-0)) line \citep{Emonts+23}. 
Indications of extended ($\gtrsim 10$ kpc at $z\sim2$, or a few times more extended than the UV emission at $z\sim5$) CO transition and \cii emission (also referred to as ``\cii halos" or ``\cii nebulae'') are reported around individual quasars from $z\sim2$ (\citealt{Cicone+21, Li2023, Jones2023a, Scholtz+23}) to the epoch of reionisation $z>6$ \citep{Maiolino+12, Cicone+15, Meyer2022}\footnote{This is nevertheless debated observationally concerning the extent of the emission and the existence of it, see, e.g., \citealt{Novak2020} from stacking analyses}, dusty star-forming galaxies at $z\sim2$ (\citealt{Gullberg2018, Rybak2019, Rybak2020, Solimano2024}) and even typical star-forming galaxies at $z\sim6$ \citep{Ginolfi+20b, Fujimoto+20, Lambert2023}, often relying on stacking techniques \citep{Fujimoto2019, Bischetti2019b, Ginolfi+20a, Fudamoto+22}. 
Also, there is even a hint of a tail that could have been stripped away from the brightest cluster galaxy at $z=1.7$ (\citealt{Webb+17, Castignani+20a}). 
Deep VLA imaging also reveals extended ($\geq$50 kpc) cold-gas reservoirs around $z=2-5$ massive galaxies \citep{Frias2023, Stanley2023}.

In the following, we take a few examples of the detection to highlight the need for AtLAST.

~~At $z=2.2197$, cid\_346 is a luminous ($\rm \log L_{AGN} [erg/s]=46.66$) AGN studied as part of the SUPER survey  \citep{Kakkad+20, Circosta+18, Circosta+21}, hosted by a massive (M$_*=10^{11}$~M$_{\odot}$) star-forming (SFR = $360$ \My) galaxy that is undergoing an explosive feedback phase.
ALMA and ACA CO~(3--2) observations by \cite{Cicone+21} reveal a giant molecular halo extending up to $r\sim200$~kpc from the AGN (the total angular diameter size of the halo is at least $50''$), which is currently the most extended molecular CGM ever mapped. 
The molecular CGM gas mass could be as high as $\sim 1.7\times 10^{12}$~M$_{\odot}$. 
However, such observations are extremely challenging for interferometers, and small changes in the imaging parameters can affect the results. Indeed, although the giant halo of cid\_346 was detected in both ALMA and ACA data by \cite{Cicone+21}, a reanalysis of the ACA data presented by \cite{Jones2023b}, who adopt a different methodology, could not retrieve the same spatial extent of the halo.
A deeper follow-up of such extended halo, in CO and \ci\, lines, would be straightforward yet game-changing with a facility like AtLAST (Section~\ref{sec:atlast}). 

~~A nearly 70 kpc-sized halo of cold molecular gas is seen in CO~(1--0) towards the Spiderweb $z\sim2$ protocluster (\citealt{Emonts2016}, Figure~\ref{fig:currentobs} (b)). 
The \HII\  gas mass of $\rm M_{H2}\sim 10^{11} M_{\odot}$ amounts to $\sim 30\%$ of its CGM mass. 
Besides being one of the very few high-$z$ systems where cold \HII\ is found in its halo, the Spiderweb is also one of the few sources in which the physical properties of the molecular CGM are studied in multiple tracers. Thanks to the wider tuning range of the SEPIA Band~9 receiver on APEX compared to the ALMA Band~9 receiver, \cite{DeBreuck+22} could detect the \cii~emission line at 601.8~GHz in this source, and spectrally deblend the ISM from the CGM component.
\textit{Hubble Space Telescope (HST)} deep imaging reveals diffuse blue light from young stars across the halo, on the same scales as CO(1-0) emission.
\textit{ALMA} observations of the molecular CGM in CO~(4--3) and \ci $^{3}P_1-^{3}P_0$ lines \citep{Emonts+18} suggest excitation conditions and carbon abundance of the molecular CGM  similar to those in the ISM of starburst galaxies. 
Deeper observations of these components with multi-phase probes, better surface-brightness sensitivity to trace the full extent of the emission from each probe, and comparison with local cluster galaxies (Section \ref{sec:multiphase}) will be the next initiative to understand galaxy evolution and formation in dense environments with AtLAST.

~~Beyond the Cosmic Noon ($z\gtrsim2$), at least two $z\sim4$ protoclusters, SPT2349 \citep{Hill+20} and DRC \citep{Oteo+18} show curious starburst coordination among galaxy members separated by hundreds of kpc. 
This coordination, along with very short gas-depletion timescales of $\tau_{\rm dep} \sim 100$ Myr within individual galaxies, hints at a much larger gas reservoir in these clusters, one that is yet to be detected. 
Indeed, if we consider the CO-based \HII\ gas mass measurements within the galaxies of these protoclusters as a more or less complete inventory, it raises a question of how such an unlikely starburst-event coordination across hundreds of kpc could happen only at the $\sim$ 6\% of the age of the Universe.
A reasonable explanation would be the presence of much more gas in the CGM and intracluster medium (ICM), whose flows (perhaps linked to the larger cosmic web) coordinate and sustain these SFRs across cluster members over much longer timescales. 
A widely distributed CO-poor gas phase would remain undetected by all past CO observations of these two protoclusters. 
Existing ALMA 12m-array observations of the \cii\ line for the SPT2349 cluster \citep{Hill+20} and the \ci\ line for the DRC \citep{Oteo+18}, do not have the uv-coverage to be sensitive to CGM-scale ($>$ a few hundreds of kpc, or tens of arcsec) gas due to the lack of short baselines.

~~The direct imaging of extended emission for `normal' star-forming galaxies is yet to be unveiled with more sensitive observations. 
Nonetheless, stacking analyses already hinted that distant galaxies are surrounded by extended cold gas haloes, likely enriched by galaxy interactions and outflows (\citealt{Fujimoto2019, Bischetti2019b, Ginolfi+20a, Fudamoto+22}). 
While \citet{Pizzati2023} attributed the observed extended emission to outflows based on simulations, current observational constraints of normal star-forming galaxies cannot prove any of those scenarios because of poor sensitivity; the emission is only revealed by \cii\, whose origin is unclear.
Whether the extended \cii's driving mechanisms are the same as those observed in (low-$J$) CO transitions is not also understood yet.
We need much more sensitive, multi-phase probes as well as direct imaging to uncover the physical properties of the CGM.\\

~~Taking all these together, current observational constraints provide an ample indication of gas reservoirs outside galactic discs that we could still be largely missing at all cosmic distances.
We may have just started scratching the surface of the cold CGM with the current facilities and AtLAST will shed light on this missing piece of information.
In the following sections, we show more quantitative assessment based on simulation results.\\

\section{Simulations: the struggle of modelling cold gas on CGM and IGM scales}\label{sec:simulations}
~~Over the last decade, cosmological hydrodynamical simulations have become increasingly better at reproducing the properties of galaxies across cosmic times such as their stellar mass, size, gas content, or metallicity (e.g., \texttt{EAGLE} \citealt{Schaye2015, Crain2015}, \texttt{IllustrisTNG} \citealt{Pillepich2018, Nelson2018} and \texttt{SIMBA} \citealt{Dave2019} simulations, see for a recent review \citealt{Crain2023}). 
All these simulations use vastly different recipes for physical processes such as stellar and black-hole feedback and it is thus remarkable that they reproduce the properties of galaxies about equally well. 
As discussed below, the properties of the cold CGM in simulations are still largely unconstrained, and thus observational comparisons are very timely.
In this section, we first describe the challenges in many current state-of-the-art cosmological simulations concerning the cold gas in the CGM (Section~\ref{sec:stateoftheart}) and provide a brief introduction to the work done by \citealt{Schimek24} (Section~\ref{sec:ponos_intro}), a cosmological zoom-in simulation focusing on sub-mm fine-structure lines aimed to bridge the current gap between simulations and the expected outcome of AtLAST (Section~\ref{sec:atlast}).

\subsection{The state-of-the-art: overlooked cold CGM and difficulties}\label{sec:stateoftheart}
~~In the last few years, theoretical emphasis around cosmological simulations has been given towards predictions of tracers of warm gas in the CGM, in particular through predictions for ionisation lines (e.g., \citealt{Ford2014, Liang2016, Nelson2018, Oppenheimer2018, Pallottini2022}). 
Modelling the CGM of galaxies requires tracking physical processes that act on cosmological volumes and timescales: cosmic accretion through filamentary streams, cooling processes within dark matter halos, star formation, accretion onto supermassive black holes, radiative and mechanical feedback mechanisms that generate galactic outflows and fountains, galaxy mergers and interactions and the consequent dislocation of gas. 
Therefore, only a zoom-in cosmological simulation that follows the evolution of galaxies across cosmic times can inform us about the properties of the large-scale ISM and CGM material, and how they evolve. 

\subsubsection{The challenges}
~~The cold gas component of the CGM in cosmological simulations has received less attention compared to a warmer, ionised gas.
Several studies demonstrate that increasing the numerical resolution of the simulation on CGM scales yields higher amounts of cold and dense material and more sub-structures
(\citealt{Scannapieco2015, Schneider2017, Mandelker2018, McCourt2018, Hummels2019, Suresh2019, vandeVoort2019, Sparre2019, Nelson2021, Lupi2022}).
For example, \citet{vandeVoort2019} use a zoom-in simulation of a Milky Way progenitor to study the CGM at an unprecedented (at that time) spatial resolution of 1 kpc (achieved through an increased mass refinement in the CGM around the simulated galaxy) and find that the radial profile of atomic hydrogen in the CGM at radii beyond 40 kpc is enhanced drastically compared to lower-resolution simulations. 
The covering fraction of Lyman-limit systems within 150 kpc from the central galaxy is almost double compared to simulations with standard mass refinement techniques. 
Similar conclusions are reached by \citet{Ramesh2023}, who use the same IllustrisTNG50 galaxy formation model to study the CGM around 8 Milky Way progenitors with an increased mass resolution within the CGM. 
These studies demonstrate that robustly modelling the CGM of galaxies and resolving the cold gas within the CGM and its interplay with galaxies through in- and outflows, requires high mass-resolution simulations. 

~~This immediately poses a challenge for cosmological simulations, because one can not computationally afford to run a full cosmological volume at the mass resolution needed to properly resolve cold gas properties within the CGM. 
The use of cosmological zoom-in simulations focusing on the CGM properties around galaxies remains the most computationally efficient way forward. 

~~Despite the recent successes in the modelling of cold gas around galaxies, there are still obvious gaps in our knowledge. 
For example, \citet{vandeVoort2019} over-reproduce the column density of \hi\, at 500 pc spatial refinement, compared to the observational result, raising a question on the simplified self-shielding model.
Similarly, while simulations of extended \cii emission at high redshift successfully predict its extent being $\sim 2$ times more extended than dust and/or \oiii emission in both analytical models (e.g. \citealt{Pizzati2020, Pizzati2023}) and cosmological zoom-in simulations (\citealt{Arata20, Katz2022, Pallottini2022}), the latter typically predict a much sharper decrease in the emission at large scales compared to observations. Since these simulations rely on different set-ups (e.g. GADGET using Smoothed
Particle Hydrodynamics, vs. RAMES using Adaptive Mesh Refinement) and physical models for both stellar feedback and FIR lines emission, the common struggle in reproducing the observational evidence demonstrates a challenge in the state-of-the-art simulations. 
Further, in their attempts to study the CGM gas around galaxies, both \citet{vandeVoort2019} and \citet{Ramesh2023} focus on Milky Way progenitors, missing the broad variety of galaxies across cosmic time. 
As discussed, a full cosmological volume is too computationally expensive to model at high resolution, but a better sampling of the rich diversity of galaxies and their environments, and testing different feedback prescriptions will be necessary (including ranges in halo masses, and various dynamical processes such as AGN outflows and mergers). 
Additionally, so far, theoretical work on the cold gas in the CGM around galaxies mostly focuses on atomic hydrogen (with a temperature of $10^4$ K or warmer), ignoring the potential presence of even colder, for example molecular, gas, and its emission. 
This is largely a technical limitation, as cosmological simulations often do not have the mass resolution nor the chemistry included to cool the gas down to 10s of Kelvin and form molecules (though see for example \citealt{Maio2022}). 

~~The inclusion of time-dependent atomic and molecular non-equilibrium chemistry within zoom-in simulations that resolve the CGM around galaxies at high mass resolution will thus mark a true step forward in the characterisation of both the atomic and molecular cold gas around galaxies, beyond the current state-of-the-art. 
Additional improvement lies in the characterisation of the influence that for example magnetic fields and cosmic rays have on the cold properties of the CGM around galaxies (e.g., \citealt{vandeVoort2021, Ponnada2022, Heesen2023, Butsky2023, Rey23, Wissing+22, Wissing+23}) that enable the survival of cold clumps and streams on large scales and thus create a multi-phase CGM.

~~For simulations to be properly predictive for observers, and to test them directly against observations, one can produce synthetic emission lines either by directly accounting for cooling and heating \citep[e.g.][]{Lupi2020, Arata20} or by using photoionisation-codes like CLOUDY \citep[e.g.][]{Vallini15, Olsen15, Olsen17, Pallottini2019, Katz2019, Vallini21, Pallottini2022, Katz2022, Schimek24}. 
For such emission line modelling an extended cold CGM component can only be recovered and predicted only if the numerical resolution beyond the central disc of the simulated galaxy is as high as that needed to model the cold ISM. 
Higher resolutions (and proper treatment of the FUV background) can lead to higher predicted extended emissions in cold gas tracers that a telescope like AtLAST could detect. We introduce an effort to such in the following section.

\subsection{A theoretical effort towards AtLAST observations: introduction of the work by \cite{Schimek24}}\label{sec:ponos_intro}

\begin{figure*}[tbh] 
\centering
   \includegraphics[width=\textwidth]{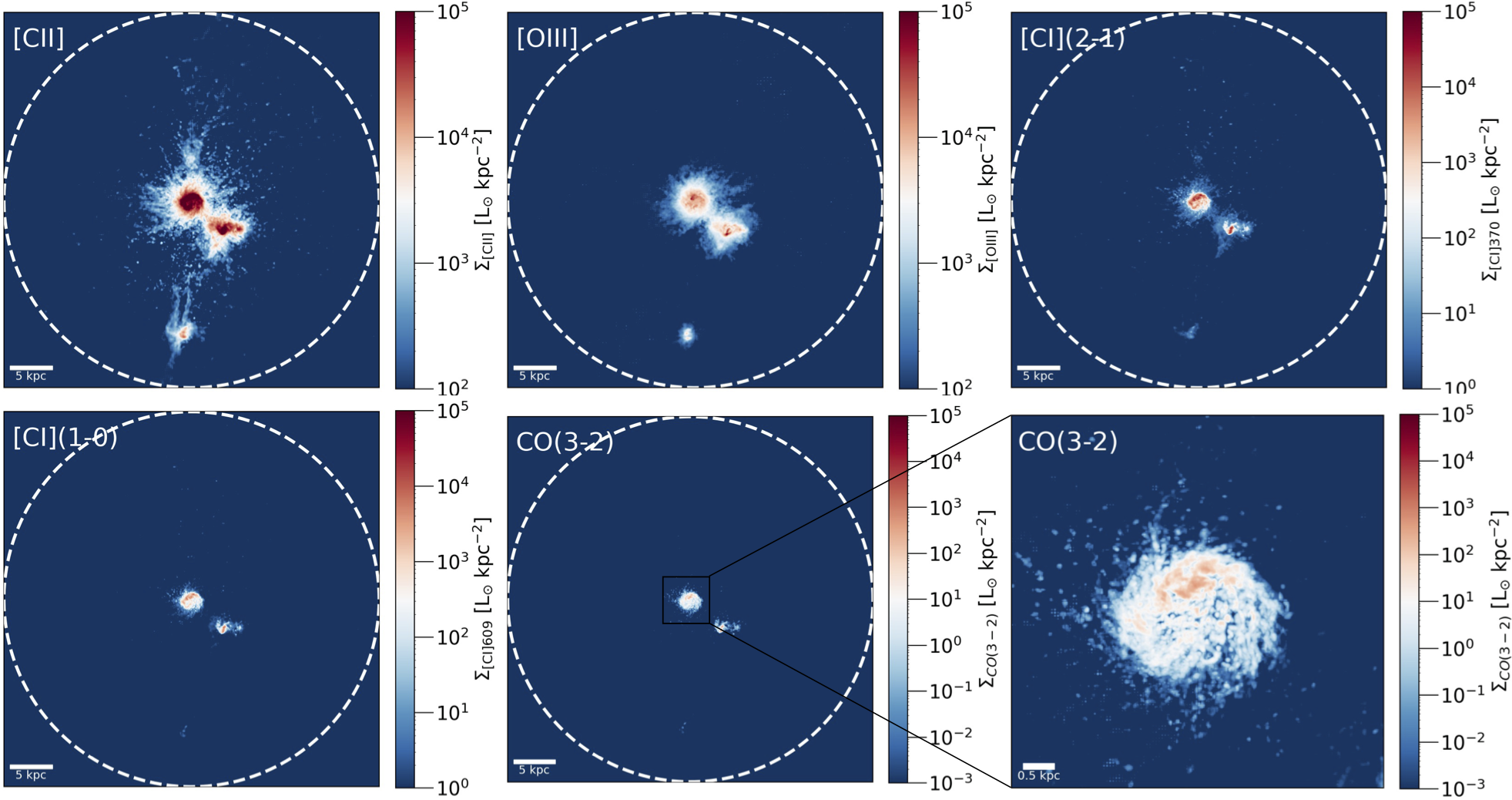}
      \caption{Simulated far-infrared and sub-mm line emission maps of the \texttt{Ponos} galaxy system at $z=6.5$, obtained using the fiducial Radiative Transfer model presented in \cite{Schimek24}. The lines shown, from the top-left to the bottom-right panels, are: \cii$158~\mu$m (a tracer of atomic, molecular and ionised gas), \oiii$88~\mu$m (a tracer of ionised gas), two \ci\, transitions (\ci~$370~\mu$m,
      \ci~$609~\mu$m; tracers of atomic and molecular gas), and CO(3--2)~$867~\mu$m, which traces molecular gas. The white circles mark the virial radius of the halo ($R_{\rm vir}=21.2$~kpc). The maps shown here are the same as Figure~11 in \cite{Schimek24} but use a different colour scheme that enhances the low-level emission. The corresponding mock spectra are shown in Figure~18 of \cite{Schimek24}.}
    \label{fig:Ponos_RTmodel}
\end{figure*}

\begin{figure}[tbh] 
\centering
   \includegraphics[width=\columnwidth]{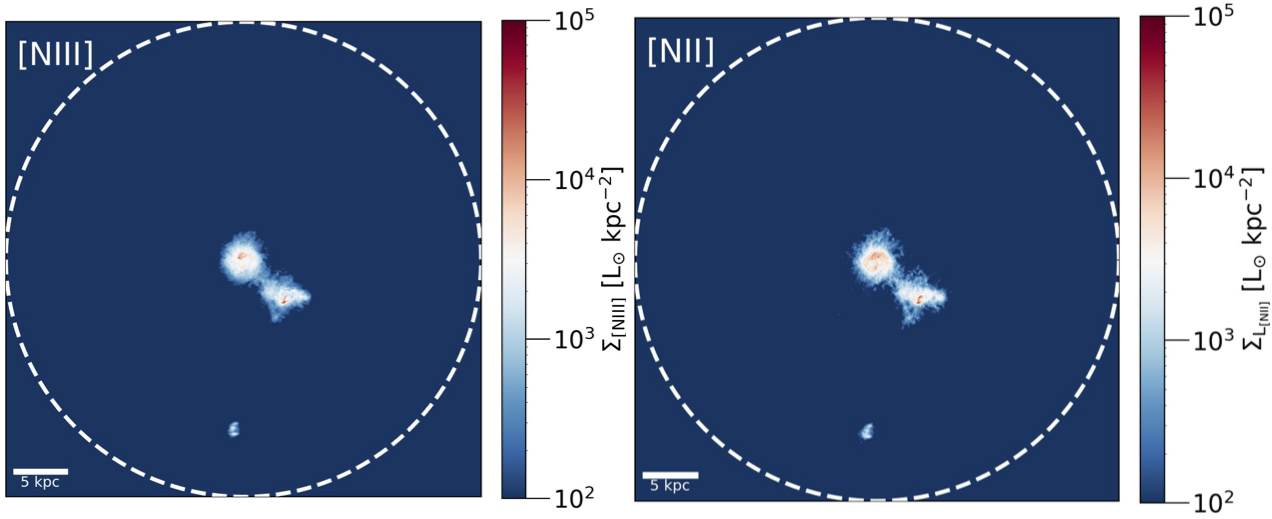}
      \caption{Simulated \niii~$57~\mu$m (left) and \nii~$\lambda 122~\mu$m (right) emission line maps of the \texttt{Ponos} galaxy system at $z=6.5$, obtained using the same fiducial radiative transfer model as in \cite{Schimek24}. These tracers are explored in a forthcoming publication (Schimek et al. in prep).}
    \label{fig:Ponos_RTmodel_Nitrogen}
\end{figure}

~~Conducted as part of the AtLAST design study, \cite{Schimek24} model prominent FIR and sub-mm emission lines for the cosmological zoom-in simulation \texttt{Ponos} \citep{Fiacconi17}.
The simulated galaxy represents a typical star-forming system at $z=6.5$ which will evolve into a massive galaxy at $z\sim0$. 
The system consists of a central disc galaxy, two merging companions showing signs of tidal interactions, and it is fed by an accreting cold gas filament, resulting in a highly multi-phase CGM.
The central galaxy has a stellar mass of $M_{*}=2\times 10^9$~M$_{\odot}$, a SFR=20~M$_{\odot}~\rm yr^{-1}$ and a virial radius of 21.18~kpc, which we assume to be the CGM radius. The particle gas mass resolution of m$_{\rm gas} = 883.4 $~M$_{\odot}$ allows for a maximum spatial resolution of 3.6~pc in the synthetic maps of the disc gas emission, and between 3.6~pc and 200~pc on CGM scales, depending on the density of the structures (the resolution is adapted to be higher in higher density structures). Such high resolution allows for the modelling of cold gas tracers, even within the CGM.

~~Synthetic maps of FIR and sub-mm line tracers are obtained using radiative transfer (RT) post-processing and the photoionisation code CLOUDY \citep{Cloudy} and are shown in Figure~\ref{fig:Ponos_RTmodel}. 
We refer to the paper \citep{Schimek24} for a detailed description of the methodology and a thorough comparison with other studies in the literature.
For this case study, the most relevant results of the \cite{Schimek24} study are:
\begin{enumerate}
    \item The \cii~$\lambda 158~\mu$m emission line is an excellent tracer of diffuse cold atomic CGM gas. Indeed, about 10\% of its total emission from the galaxy system originates from cold $T<10^4~K$ gas residing in diffuse CGM components that are not ascribable to ISM material (of either the main disc or the companions), such as the cosmic filament and the tails and bridges connected to the merging process. Unfortunately, due to the multi-phase nature of \cii emission, having such a line alone without additional tracers, cannot provide solid constraints on the physical properties of the CGM gas at high z. 
    \item At least $\sim20$~\% of the total \oiii~$88~\mu$m line emission arises from a puffy halo surrounding the main disc, ionised by newly formed stars and possibly linked to feedback from supernovae, hence it traces different CGM components from \cii.
\end{enumerate}

~~In a forthcoming paper (Schimek et al., in prep), the same radiative transfer (RT) modelling methods are applied to study additional FIR tracers, such as the \niii~$57~\mu$m and \nii~$\lambda 122~\mu$m lines, whose synthetic maps are shown in Figure~\ref{fig:Ponos_RTmodel_Nitrogen}. These lines are complementary to \cii and \oiii and can provide useful constraints through line ratios, but preliminary results do not indicate that such tracers can be particularly bright on CGM scales.

~~On the other hand, the results of \cite{Schimek24} should be considered conservative estimates (lower limits) for the cold and dense gas mass in the CGM.
The reasons are as follows. (i) First of all, the \texttt{Ponos} simulation does not include AGN feedback, which, according to simulations \citep{Costa+22}, is crucial to produce a gas-rich CGM in the early universe through powerful outflows.
(ii) Secondly, despite its unprecedented resolution (for a cosmological zoom-in simulation), the emission from molecular gas tracers such as CO is still underestimated by the model. 
This is not the case for \cii and \oiii, whose resulting total luminosities show a general good agreement with observational values, although the simulation is still struggling to retrieve the high \oiii/\cii ratios seen in high-z observations (but there are promising results from a future follow-up work by Nyhagen et al. (in prep)). 
(iii) Finally, the simulation allows disentangling completely the disc material of the central galaxy and its merging companions from genuine diffuse CGM components, while this is not often the case for observations. 
Hypothetical observations of the \texttt{Ponos} system at arcsec resolution would account for the southern and western satellites visible in the maps in Figure~\ref{fig:Ponos_RTmodel} as part of the CGM, while they are not considered in the CGM emission fractions discussed above.

\section{What AtLAST can do}\label{sec:atlast}

\subsection{Observing the CGM of a typical star-forming galaxy system at $z\sim6$}\label{sec:sec41}

\begin{figure*}[tbh] 
\centering
   \includegraphics[scale=.45]{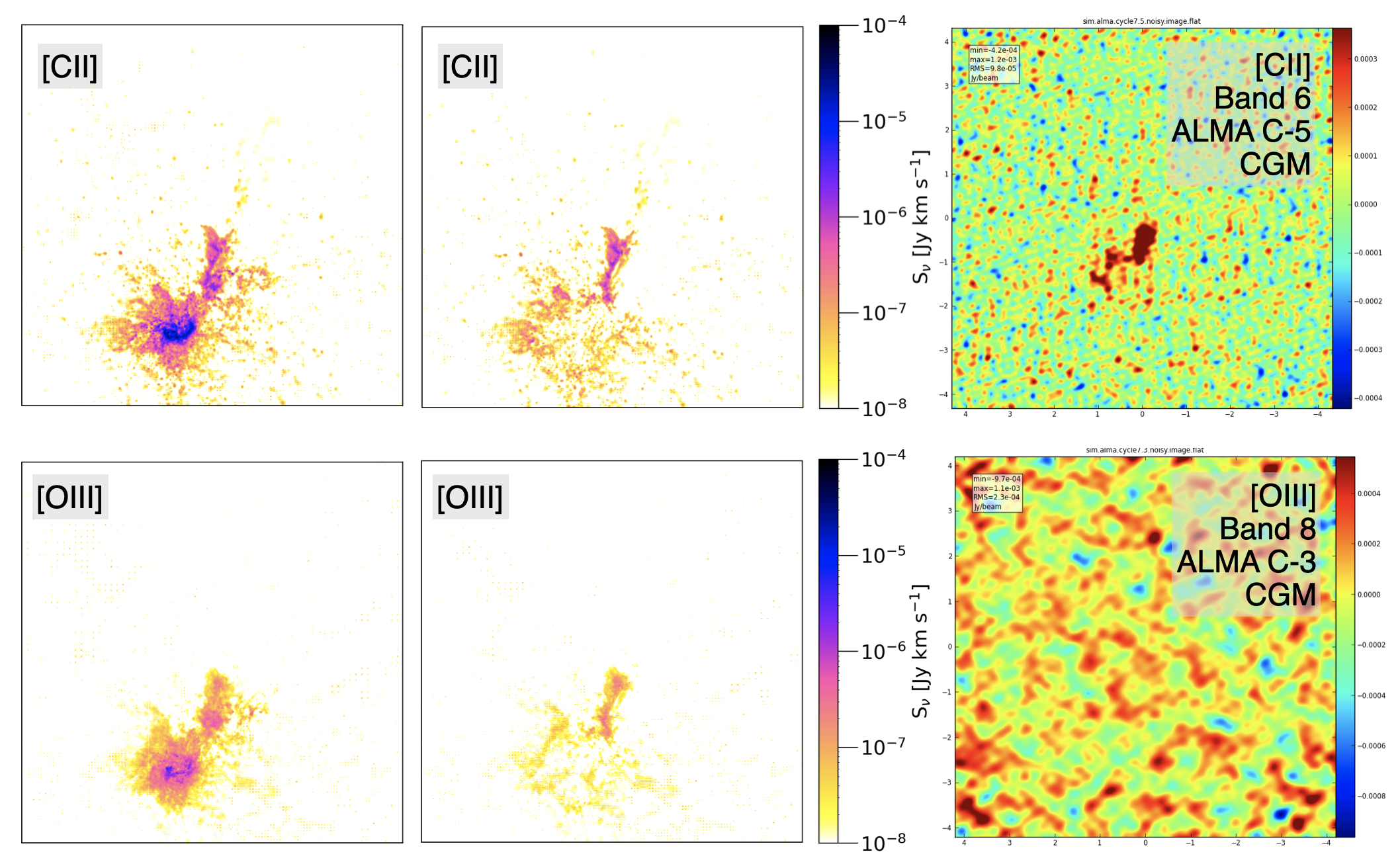}
      \caption{ALMA mock observations of the \cii (top) and \oiii (bottom) line emissions from the CGM of the \texttt{Ponos} galaxy system, obtained with the CASA simulator \citep{CASA}. The field of view 
      corresponds to the central 25 kpc region of the simulated galaxy halo (4.4 arcsec at z=6.5).
      The left panels show the total line emissions within the central 50~km~s$^{-1}$ channel. The middle panels show the same map after subtracting the ISM components, corresponding to the CGM-only contribution, which was given as input to the CASA simulator. The right panels show the resulting CASA mocks obtained with integration time set to 10 hours on-source (without overheads). We note the change in orientation of the maps shown in Figure~\ref{fig:Ponos_RTmodel}. This figure is a modified version of that shown in \cite{Schimek24}.}
   \label{fig:CASA_mocks_z6.5}
\end{figure*}

\begin{table*}[tbh]         
\centering
    \begin{tabular}{lccc}        
    \hline\hline
    Target & AtLAST    & ALMA C-1  & ALMA C-5    \\ 
    \hline 
    \cii at 253.4~GHz (S/N=10 on $\Delta v=50$~km~s$^{-1}$ channels)  & 9.7 hours      & 2 days      & 3555 days   \\
    \oiii at 460 GHz$^{\dag}$ (S/N=3 on $\Delta v=100$~km~s$^{-1}$ channels)    & 2.3 days       & 211 days  &       \\  
    \hline 
    \end{tabular} 
    \caption{On-source time estimates for ALMA and AtLAST observations of the (red-shifted) \cii (for S/N=10) and \oiii (for S/N=3) line emissions from the CGM based on the simulated star-forming galaxy halo at $z=6.5$, from \cite{Schimek24}. $^{\dag}$ For \oiii, a central frequency of 460~GHz instead of 452.4~GHz (the true frequency of \oiii at z = 6.5) was used to improve atmospheric transmission and a PWV=0.5~mm is assumed. The comparisons of the on-source times show that the higher surface brightness sensitivity provided by AtLAST in the higher frequency bands can allow for detection of the \oiii emission from the CGM of a high-z galaxy 100 more efficiently than with ALMA, even for relatively compact high-z sources.}\label{table:Ponos}
\end{table*}

~Here we use the theoretical predictions of \cite{Schimek24} to compare the performances of AtLAST and ALMA in observing the CGM component of the simulated \texttt{Ponos} galaxy system, which is representative of the $z\sim6$ star-forming galaxy population. 
At $z\sim6.5$ the virial halo of \texttt{Ponos} has a projected angular diameter of $7.5$~arcsec, hence it is a fairly compact source that can be easily captured even by the small FoV of ALMA. 
We focus on the central 25~kpc of the system (4.4~arcsec).

~~Figure~\ref{fig:CASA_mocks_z6.5} shows the simulated two brightest tracers of the CGM: \cii and \oiii. We produce the mock images where we consider only the CGM emission from the system, obtained after subtracting the ISM of the main galaxy and its companions (as explained in \cite{Schimek24}, Figure~\ref{fig:CASA_mocks_z6.5} middle). 
Indeed, while an interferometer, thanks to its high point source sensitivity and spatial resolution, is certainly more suitable than AtLAST for resolving the ISM of high-z galaxies, 
it struggles to detect the diffuse CGM, even from a compact source such as the one considered here (Figure~\ref{fig:CASA_mocks_z6.5} right). 
The reason for this is that the CGM emission is extended and diffuse; the sensitivity per beam required to detect such a component scales down by the number of (synthesised) beams contained in it, i.e. by $(D_{source}/D_{beam})^2$, dubbed as `surface brightness dimming' in \cite{Carniani2020}. 
In addition to this, an interferometer is based on the principle of wave interference to detect a signal and produce an image. 
For emission that is more extended than the so-called `largest angular scale’, which is set by the length of the shortest baseline, the interference-fringes overlap and cancel each other out. 
This means that an interferometer will never detect gas reservoirs on scales that are more extended than those corresponding to its minimum baseline and the loss of large-scale flux can be very significant already on much smaller scales (see \citealt{Plunkett+23}). In comparison, for a single-dish such as AtLAST, the maximum detectable angular scale corresponds to the maximum field of view covered by detectors (some flux loss at larger scales is expected due to scanning).

~~Assuming the peak CGM flux density values derived by the RT modelling (3.8~mJy and 1.34~mJy respectively for the \cii and \oiii lines), the ALMA Cycle~10 observing tool and the current version of the AtLAST sensitivity calculator\footnote{\href{https://www.atlast.uio.no/sensitivity-calculator/}{atlast.uio.no/sensitivity-calculator/}} deliver the integration times summarized in Table~\ref{table:Ponos}. 
The goal sensitivities and channel sizes are also reported in the table. For ALMA, the table shows values obtained with the most compact configuration (ALMA C-1) and with an intermediate configuration that is commonly adopted to observe the ISM of high redshift galaxies (ALMA C-5) to achieve S/N=10 for \cii detection. Table~\ref{table:Ponos} clearly shows that, while AtLAST can detect the CGM of a typical $z\sim6$ galaxy system in both \cii and \oiii (for the latter, a deep observation is needed), this is an impossible task for ALMA.
However, deep integrations with ALMA may still be able to detect high-surface brightness details of the CGM emission, as shown in the CASA mock observations reported in Figure~\ref{fig:CASA_mocks_z6.5}, obtained with the CASA simulator by \cite{Schimek24}. 

\subsection{Observing the molecular CGM at $z\sim0$}

\begin{figure*}[tbh] 
\centering
   \includegraphics[width=0.9\textwidth]{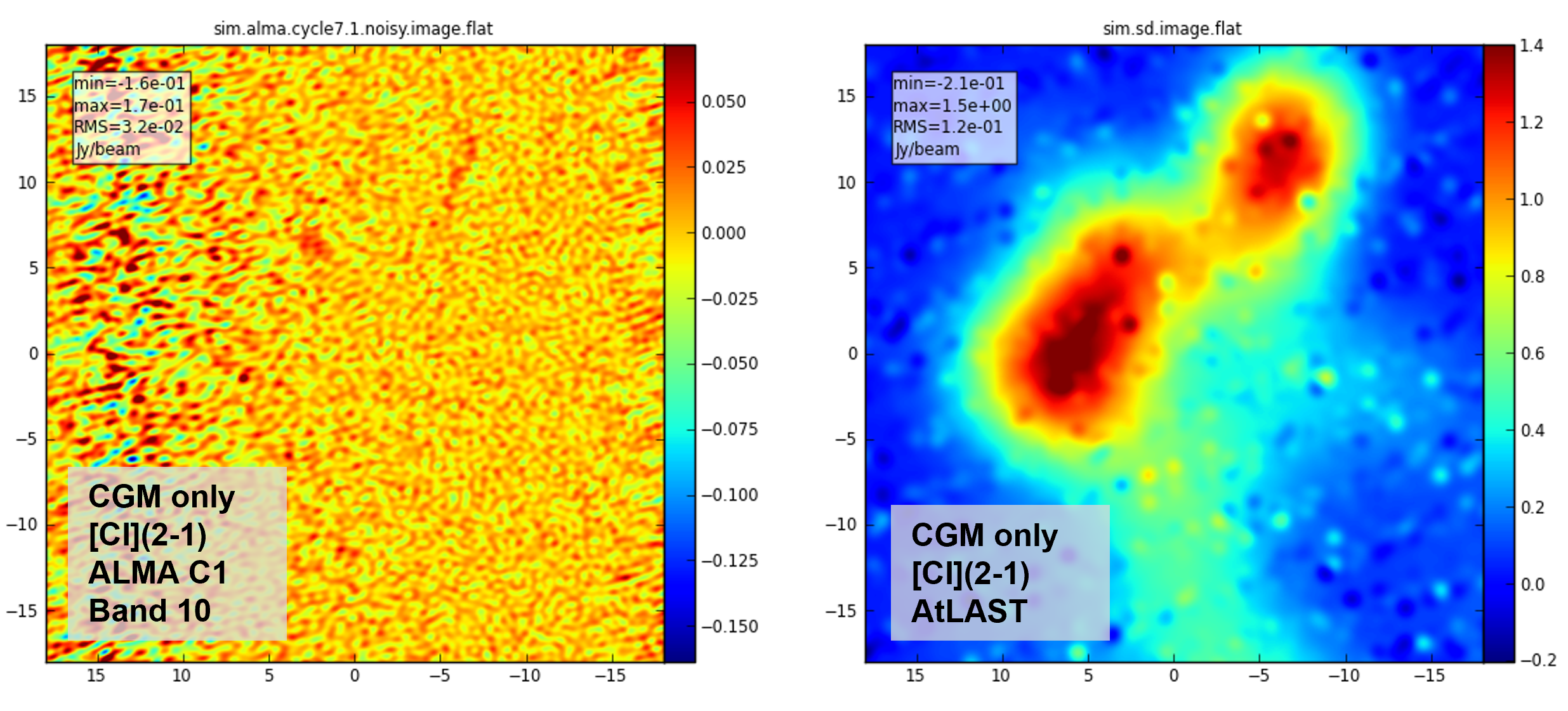}
      \caption{Mock observations of the \ci line emission from the CGM of a \texttt{Ponos}-like galaxy at $z = 0.01$, obtained with the CASA simulator \citep{CASA}. The morphology of the input image is similar to that of the source shown in the central panels of Fig~\ref{fig:CASA_mocks_z6.5}, with the difference that the system has been artificially shifted to $z=0.01$, and so has a larger angular size (while keeping the same physical size). The observed frequency is 801.326~GHz, which falls within the ALMA Band 10 range. The left panel shows the results obtained with ALMA in configuration C-1 (synthesised beam $0.53''\times0.38''$, pixel size is $0.05''$), with a 10-hour on-source time. The right panel shows a simplified AtLAST mock observation (angular resolution of $1.96''$) produced by tweaking the CASA simulator to mock a 50-m single dish and using a 10-hour on-source time per beam to mimic a multi-beam heterodyne receiver.}
   \label{fig:CASA_mocks_z0}
\end{figure*}

~~Because of computational limitations, cosmological simulations with such a high numerical resolution as \texttt{Ponos} cannot be evolved down to $z\sim0$. 
As an exercise to demonstrate the capabilities of AtLAST to detect the molecular CGM in local galaxies, we artificially shifted the \texttt{Ponos} simulation to $z=0.01$. Figure~\ref{fig:CASA_mocks_z0} reveals the recovery of the \citwo (hereafter, \ci(2-1)) line emission in the CGM component\footnote{The morphology of the CGM is the same as shown in Figure~\ref{fig:CASA_mocks_z6.5} in \cii and \oiii.} after 10-hours on source with ALMA and AtLAST, under the hypothesis that the simulated galaxy has $z=0.01$. 
The figure demonstrates clearly that molecular CGM observations with ALMA, especially using high-frequency tracers such as the \ci~(2-1) line, are extremely challenging for local galaxies due to their larger angular sizes, even for a relatively compact source such as the simulated \texttt{Ponos} galaxy, and even when the most compact array configuration is used. Our simplified AtLAST mock\footnote{We note that a proper AtLAST mock tool for multi-beam heterodyne observations has not been developed yet, while a continuum mock tool has been recently released by \cite{vanMarrewijk2024}.} shows promising results for this science goal. 

\section{Technical Requirements for AtLAST}\label{sec:tech_req}
~~Pilot studies done with other telescopes (APEX, ALMA, ACA, ATCA, etc, see summary in Section~\ref{sec:obs_constr}), despite being severely affected by technical limitations, show a clear signature of very extended cold atomic and molecular gas emission around galaxies at all cosmic epochs, which AtLAST could study.
The technical requirements needed to achieve the scientific goals of the CGM science case are detailed below and summarised in Table~\ref{tab:summary}. \\

\noindent{\bf (I) Need for a 50-m single dish}

~~The CGM investigation requires a large aperture single dish as opposed to an interferometer due to the nature of the low surface brightness emission, which is extended and diffuse, at any redshift (Figure~\ref{fig:CASA_mocks_z6.5}) but especially for nearby galaxies (Figure~\ref{fig:CASA_mocks_z0}). 
Nyquist-sampled single-dish maps are indispensable for "filling in" the central (u,v) hole in the Fourier plane that characterises interferometric measurements. 
Without the addition of single-dish observations, there is the danger of missing up to 70\% of the flux or more \citep{Plunkett+23}, and the missing flux affects spectroscopy as well as imaging on scales as small as a few times the synthesised beam \citep{Hacar+18}.
The loss of large angular scales in interferometric observations is an obstacle to the detection of the faintest and most diffuse components of the cold CGM, i.e. the sub-mm counterparts of the $\sim100$~kpc - long streams and filaments revealed by HI observations (see Section~\ref{sec:obs_constr}, and \citealt{Lucero+15, deBlok+18}) and of the large Ly$\alpha$ nebulae that can now be easily studied with optical telescopes equipped with integral field units (e.g. \citealt{Borisova2016, Cai+19}) 

~~The small synthesised beams of interferometers, especially at sub-mm frequencies, lose the total signal (the flux) and break it into too many resolution elements, or even resolve it out completely, hence impeding the detection of diffuse CGM. The cold CGM, even when detected (at low S/N) thanks to few dedicated pilot studies conducted with the currently available interferometers, often appears as disconnected low S/N `blobs' (as a result of the small Gaussian beam and sparse filled Fourier plane), whose interpretation can be uncertain hence slowing down scientific progress. 
The limited angular resolution offered by a single dish is not a major concern for this science case, because the most interesting (and elusive) CGM structures are diffuse (cosmic streams, outflows, tails), and any unresolved source detected with a single dish can be later followed up with an interferometer that can access the same portion of the sky.

~~ The \textit{too good} angular resolution of interferometric facilities (without good sensitivity to diffuse emission) is not only a major concern for local galaxy studies, but it also precludes progress at high-$z$.
The exercise shown in Figure~\ref{fig:CASA_mocks_z6.5} and Table~\ref{table:Ponos} demonstrate that, for a typical star-forming galaxy system at $z\sim6$ with a compact projected CGM size of a few arcsec, ALMA would still struggle to image the CGM emission in its two most luminous tracers (\cii~158$\mu$m and \oiii~88$\mu$m). 
The ALMA \cii~ image after 10 hours of integration (Figure~\ref{fig:CASA_mocks_z6.5}) would recover only the brightest CGM features, hence missing the full picture.

~~ The large aperture of 50~m is a key requirement for CGM science not only because of the sensitivity it provides, which is essential for this science but also because it would provide the (so far, missing) fundamental complement to ALMA, enabling the much-needed overlap in short {\it uv} baselines required for a proper ALMA and single-dish data combination. Placing AtLAST close to ALMA, on the Chajnantor Plateau in Chile, would grant access to its same portion of the sky which is ideal for ALMA follow-ups of the AtLAST-detected point sources, and it would allow ALMA + SD data combination (See also Section~\ref{sec:synergy_interfero})
~~At present, the four designated single-dish telescopes of ALMA that should provide total power (TP) measurements for the 12m and/or the 7m array are inadequate for extragalactic studies because of their poor sensitivity (they cannot detect the faint features for which we need the sensitivity of ALMA) and poor performance in terms of atmospheric correction (Manolidou et al, 2023, ESO Internship Report, Priv. communication).
\\

\noindent{\bf (II) Need for a large field of view}
~~For a single-dish telescope, the maximum recoverable scale is set by the maximum portion of the field of view filled with detectors. 
As discussed in Section~\ref{sec:obs_constr}, depending on the type of sources and redshift, the extent of faint CGM can reach a few arcminutes and even a few degrees. 
Degree-scale structures are most relevant for Local Group objects and the Milky Way. 
Therefore, we recommend a FoV of 1 degree which will enable CGM science at all redshifts and comparison of the evolution of CGM structures across cosmic times and connect it with our galaxy evolution theory.\\

\begin{figure*}[tbh] 
\centering
   \includegraphics[clip=true,trim=0.2cm 1.6cm 0.2cm 3.4cm,width=.95\textwidth]{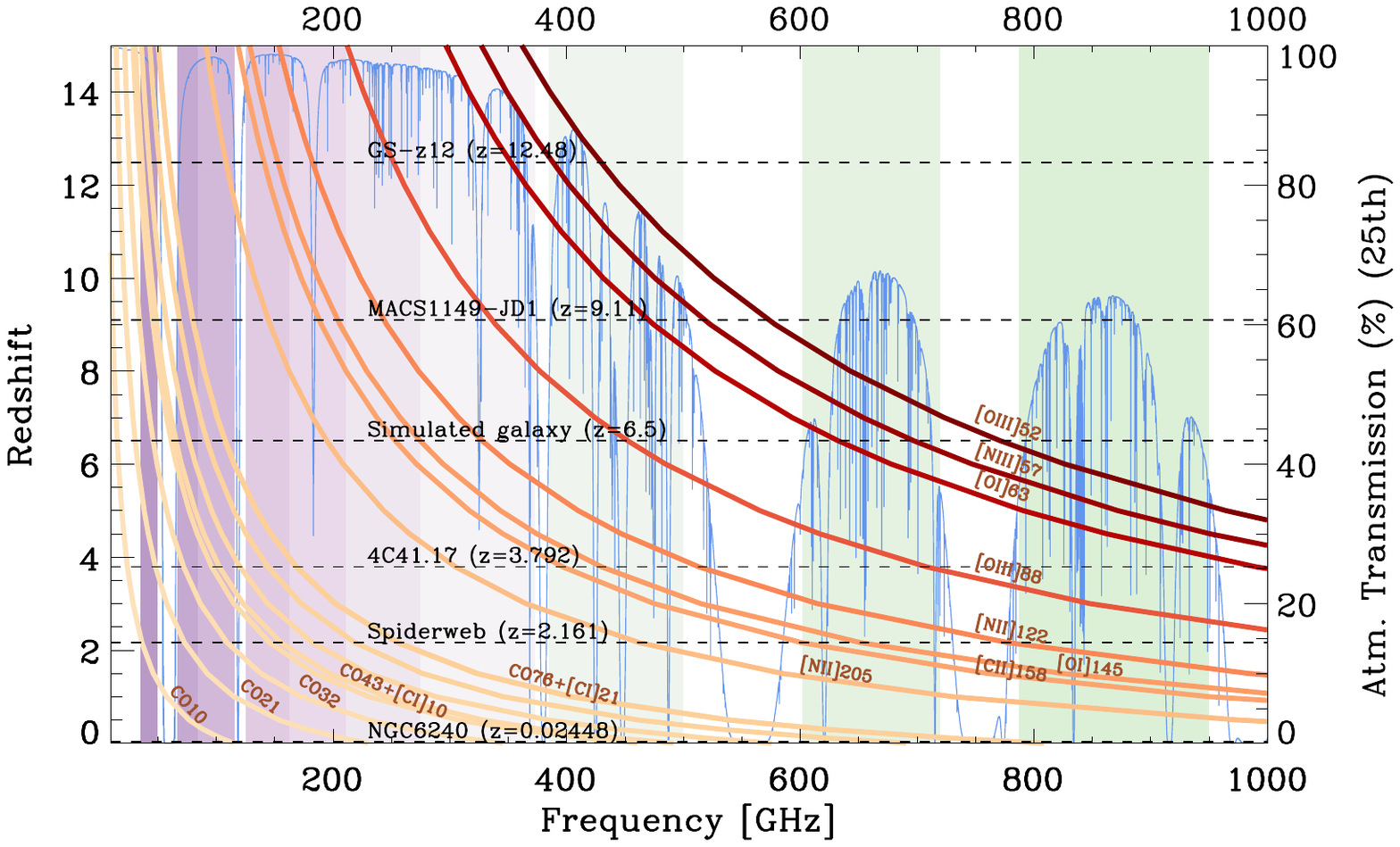}
      \caption{Evolution of the observed frequency of the brightest sub-mm and far-infrared emission lines (orange/brown lines) as they get redshifted into the sub-mm bands at different redshift values. The blue spectrum in the background shows the top quartile of the atmospheric transmission at the Chajnantor Plateau (about 5100 metres above sea level, derived using the atmospheric \texttt{am} code, \citealt{Paine2019}), where the corresponding transmission is reported on the y-axis on the right. The coloured vertical bands correspond to the ALMA bands (from Band 1 to Band 10). The horizontal dashed lines indicate the redshifts of a few sources discussed in the paper, and of two additional high-z galaxies from the literature \citep{D'Eugenio2023, Hashimoto2018}. This plot demonstrates the importance of a broad wavelength coverage for the CGM science case, and it highlights how crucial the high-frequency bands are to observe the brightest tracers of the cold CGM in galaxies at $z>1$.}
   \label{fig:plot_bands_lines}
\end{figure*}

\noindent{\bf (III) Sub-mm capabilities}

~~Being able to detect multiple tracers of the CGM in the same source is essential to study the physical properties of the gas, its density, temperature, metal content (and the CO-to-H$_2$ conversion factor), and so its origin. 
Given the wide range of redshifts (from $z=0$ to $z=10$) that is of interest for the CGM science case, a telescope with sub-mm (e.g. $\lambda < 600~\mu$m) capabilities would have crucial access to the most promising (because expected to be brightest, see Figure~\ref{fig:Ponos_RTmodel} and Figure~\ref{fig:Ponos_RTmodel_Nitrogen}) gas tracers of the cold CGM in a larger volume of the Universe, such as the atomic carbon (\ci) lines at rest-frame 609.135$\mu$m and 370.415$\mu$m, the \cii line at 158~$\mu$m, and \oiii at 88$\mu$m.
The atomic carbon lines for example can probe CO-dark H$_2$ reservoirs (e.g., \citealt{Lada1988, vanDishoeck1988, Wolfire2010}) expected to trace diffuse CGM components affected by the UV background and by Cosmic Rays. 
The accessible lines at different redshifts are shown in Figure~\ref{fig:plot_bands_lines}.
For these reasons, AtLAST needs to be located in a high dry site and needs to have a high surface accuracy to access the sub-mm window.\\

\noindent{\bf (IV) Instrumentation requirements}

~~CGM science would benefit the most when simultaneous imaging and spectroscopy capabilities, hence focal plane arrays of heterodyne detectors, are implemented on AtLAST. Concerning the limited window to observe high-frequency observations throughout the year even at high altitudes (based on the statistics of current ALMA observations), simultaneous observations at different frequency ranges would make the observations more efficient. Simultaneous multi-band observations (e.g. Band~6 and Band~7) would enable targeting multiple lines in the same long exposure, hence they would produce a substantial scientific gain for the most time-consuming CGM observations.

~~A wide simultaneous bandwidth of a minimum of 4~GHz and, preferably, 8~GHz is needed. 
CGM emission lines in extragalactic sources are expected to be broad ($\sim1000$~km~s$^{-1}$ or higher, e.g. \citealt{Cicone+14, Cicone+21}), and sampling a sufficient number of adjacent spectral channels to perform a proper baseline subtraction is essential for this science.
The stricter requirement of 8~GHz is especially important for high-frequency receivers, since at 900~GHz, a bandwidth of 4~GHz corresponds to only $\sim1300$~km~s$^{-1}$, which is not enough to cover a broad extragalactic CGM line and its adjacent continuum. 
This is particularly true for observation where also companion galaxies at different redshifts contribute to the overall signal, or where multiple lines of interest fall in the band (notably \ci (2-1) 370.415$\mu$m and CO(7-6) 371.647$\mu$m).
Wide bandwidth is also important for local sources such as the MS, for which the velocity gradient across the MS could broaden the emission lines can be shifted by several $\sim100$~km~s$^{-1}$ with respect to the systemic velocity.
We note that this particular science case does not benefit from huge bandwidths (wider than $8$~GHz), because the transitions that are bright enough for CGM studies are few and they are not clustered in frequency. 

~~ A wide tuning range within each band can be crucial for specific targets. One salient example is the case of the Spiderweb protocluster discussed in Section~2 (see also Figure~\ref{fig:plot_bands_lines}), whose particular redshift means its (bright) \cii~ emission line is outside of the allowed ALMA Band~9 frequency range, but within the allowed APEX SEPIA Band~9 frequency range \citep{DeBreuck+22}. 

~~The spectral resolution needs to be of at least 5-10~km~s$^{-1}$ for extragalactic studies. A higher spectral resolution of 0.1-0.2~km~s$^{-1}$ could benefit the study of gas dynamics in the CGM of Local Group sources, such as the MS, but is not necessary for the study of the extragalactic CGM on which this paper focuses on (see, e.g., \citealt{Cicone+21} for observational example and \citealt{Schimek24} for simulation).

~~In addition, CGM science requires stable spectral baselines. 
Because of the faint and broad nature of CGM emission lines, systematic effects due to baseline instabilities would not only conspire to create false positive detections (or false negative) in the presence of low-frequency standing waves, but they would also prevent the S/N from scaling with channel size and integration time according to the radiometer equation ($S/N\propto\sqrt{\Delta v \cdot \Delta t}$), hence nullifying the benefit of long integration and stacking analyses. 
An internal wobbler solution such as that suggested by R. Hills in a dedicated AtLAST memo\footnote{\href{https://www.atlast.uio.no/documents/memo-series/memo-public/wobbler_for_atlast.pdf}{AtLAST memo by R. Hills on the Internal Wobbler solution}} would be recommended for this science case. 
We note that, although this solution may be possible only for a smaller FoV instrument (and not for the full 1-2~degrees FoV of AtLAST), this would not be a major issue, since the stable spectral baselines are most critical for high-$z$ CGM observations, where lines are broad and one beam collects the signal from regions that correspond to several physical kpc.
We also note that, in the future, fast scanning with multi-beam receivers may be enough to deliver the flat baselines needed for this science (see \citealt{Mroczkowski2024}).\\

\noindent{\bf (V) A note on continuum observations}

~~Although this paper focuses on gas tracers, for completeness, we note that complementary continuum observations with AtLAST are needed to infer the dust content, its temperature and properties. Without this key information not only the level of star formation in the CGM of galaxies cannot be derived but also we miss the total energy budget needed to constrain the models.
Such continuum observations can also detect star formation activity from small satellites, and so help de-blend true CGM components from the ISM of small companions. Whereas line emission is kinematically resolved across channels (with typically only a fraction of the emission covering a single channel), continuum emission is not, hence the limitation of interferometers to recover widespread emission is even more severe for continuum observations.
Sub-mm polarimetry is also a promising avenue for tracking magnetic fields from ISM to CGM scales in both local and high-z galaxies \citep{LopezRodriguez2023a, Geach2023}, and we refer to the companion AtLAST case study by Liu et al. (2024, in prep.) for an in-depth discussion.\\

\noindent{\bf (VI) Operational requirements}

~~Granting observing time to individual PIs (e.g. through a proposal selection process) is crucial for this science case since it cannot be performed via general-purpose surveys as it has specific requirements, which can change for different targets and redshifts. 
Depending on the target properties (redshift, line luminosity, angular size) and the transition of choice, CGM science can be pursued through PI-driven programs that range from small projects (10 hours or less of observations) for local sources and bright sources at $z\sim2$ such as cid\_346 or the Spiderweb \citep{Emonts+18, Cicone+21}, up to large programmes of several 100s of hours, for example for a survey of the central portion of the MS ($\rm 20~deg\times100~deg$), which is the main CGM component of the Milky Way.

~~ Given the long integration times, granting the PI the possibility to interact with the observers, check the quality of the data while being taken and, in case, refine the observing strategy, can help maximise the scientific output.

~~ Calibration of the data may be performed through standard procedures, however, given the faint nature of the CGM signature, PIs may want to reanalyse the data themselves and clean the data from scans that are affected by baseline issues. A tool to check and record automatically the quality of the spectral baselines would be useful for this science.\\


~~The technical setup proposed to address this transformational science case is unique to a facility such as AtLAST, as it is unparalleled by any other planned or current facility. 
We provide the summary of the technical requirements in Table~\ref{tab:summary}.

\begin{table*}[!ht]
    \centering
    \begin{tabular}{lc}
        \hline\hline
        Parameters                       & Requirements  \\
        \hline
        Band                 &         ALMA Bands 1 to 10 \\
        Polarisation products$^{\dag}$                &  N/A  \\
        Observations                        & Mostly line, but continuum should be considered too \\ 
        Mapping type                       &  heterodyne arrays or high-spectral res IFUs \\ 
        Observing mode                    & targeted (PI-driven science) \\ 
        Central Frequency                 & redshifted frequencies of various FIR and sub-mm tracers \\
        Total Bandwidth                   &  8 GHz (contiguous) \\
        Spectral resolution               & $\lesssim 5-10$ km~s$^{-1}$$^{\ddagger}$  \\
        Number output channels           & same as native spectral resolution \\
        Angular resolution             & a few arcsec \\
        Mapped image size              & A few degrees for most targets, except for the MS ($\rm 20~deg\times100~deg$) \\
        Peak flux densities           &  $\sim$mJy/beam, depending on target \\
        I rms                         & tens of $\mu$Jy/beam  \\
        Q rms$^{\dag}$                         & N/A  \\
        U rms$^{\dag}$                       &  N/A  \\
        V rms$^{\dag}$                       & N/A  \\
        Polarised peak flux density$^{\dag}$ & N/A  \\
        Polarised fraction$^{\dag}$                 &  N/A \\
        Dynamic range                      &  high \\
        Absolute flux calibration           &  10-20\% \\
        Integration time                   &  from a few hours to 100s of hours  \\
        Additional requirements      & stable spectral baselines \\
        \hline
    \end{tabular}
     \caption{Summary of technical requirements for the CGM case study. $^{\dag}$ Because of the faintness of CGM features, we do not consider polarization products. However, SOFIA observations of local galaxies such as the Antennae merger detected polarised dust emission on scales of several kpc \citep{LopezRodriguez2023a}, hence showing that polarimetry on CGM scales is indeed a promising future avenue. We refer to the companion AtLAST case study on nearby galaxies by Liu et al. (2024, in prep.) for the polarimetry requirements. $^{\ddagger}$ For extragalactic studies. As noted in the text, a higher spectral resolution is needed for the MS and the CGM of Local Group sources.}
    \label{tab:summary}
\end{table*}

\section{Synergy with other facilities}\label{sec:synergy}
\subsection{Single-dish and interferometric data combination for a proper ISM subtraction and cold gas cycle study}\label{sec:synergy_interfero}
~~The synergy between current sub-mm interferometers and AtLAST is undeniable.
Interferometry is more sensitive to denser, compact emissions from the ISM, and so can be used to perform a proper subtraction of ISM components from any AtLAST observation where ISM and CGM are blended.
To first order, AtLAST will measure the contribution of the CGM by investigating the curve of growth and the changes in spectral line shapes with different aperture sizes. 
However, when combined AtLAST and interferometric observations are obtained, they can place solid constraints on each contribution in the emission line.
For example, \citet{Cicone+21} demonstrated how such constraints can be obtained based on the ACA and ALMA main array. 
We expect more robust measurements to be obtained with AtLAST and the existing/future interferometries.

~~ Further, when both observations are combined, as we envisaged in the introduction, we anticipate the entire baryonic cycle can be finally mapped both spatially and kinematically. 
Sensitive mapping of a few systems will guide us on how the cold gas cycle operates from 100s kpc to less than 1 kpc. 
In this regard, confirming the kinematic connection between the cold CGM and ISM will be essential and this is only possible when both observations are obtained; to some extent, we may be able to see some signature of non-circular motions from the CGM emission, which may trace the cold accretion into the galactic discs.
Therefore, there is great excitement about completing the full baryonic cycle (see also below) with a great potential of synergies with (upgraded) currently existing sub-mm interferometry. 

\subsection{Multi-phase CGM}
As described in the earlier sections, the CGM has an imprint of past accretion and feedback, making it by nature multi-phase.
Therefore, constraining the individual budget (mass and metallicity) of each gas phase (cold/warm/hot/neutral/ionised), their distribution, and understanding their interplay is the ultimate synergy with current/future facilities at different wavelengths.
In the following, we briefly describe how individual probes are interconnected and provide examples of instruments.

\subsubsection*{Warm–hot intergalactic medium (WHIM): AtLAST, X-ray missions -- XRISM, Athena, LEM}
The warm-hot intergalactic medium (WHIM) can be probed via direct detection of absorption lines in rest-frame UV/Xray and in the continuum 
of the hot CGM plasma \citep{Mathur2022, Nicastro2023} and via indirect detection exploiting the Sunyaev-Zeldovich (SZ) effect. 
While direct detection is extremely challenging to achieve at large redshift even with future facilities (because of lack of sensitivity or small field of view), the SZ effect, being redshift independent is a very powerful tool (see Di Mascolo et al. (2024, in prep.), for future role of AtLAST in this respect). 

~~The current JAXA/NASA mission, X-Ray Imaging and Spectroscopy Mission (XRISM; \citealt{Tashiro2020})\footnote{\url{https://xrism.isas.jaxa.jp/en/}} with a soft X-ray spectrometer with $5-7$~eV resolution calorimeter may combine high-resolution and large effective area and help to open a new window on CGM science. 
The X-ray Integral Field Unit (XIFU) on board Athena\citep{Barcons2012}\footnote{\url{https://www.the-athena-x-ray-observatory.eu/en}} has high spectral resolution and large area, making it well-suited for the studies of the warm-hot CGM, targeting the ${\rm O~VII\, K\alpha}$ and ${\rm K\beta}$ and the ${\rm O~VIII}$ lines, despite being not suitable for CGM studies in the nearby Universe.

~~The Line Emission Mapper (LEM;\citealt{Kraft2022})\footnote{\url{https://www.lem-observatory.org/}} concept, prepared for the NASA 2023 Astrophysics Probes call for proposals, foresees a large-area X-ray integral field unit, which will effectively map the soft X-ray line emission at a spectral resolution of  1–2 eV and make it possible to separate the faintest emission lines from the bright Milky Way foreground. The combination of effective area and the field of view solid angle will allow mapping of faint extended objects with sizes comparable to or greater than the field of view, such as nearby galaxies, clusters, and their CGMs. 

~~The possibility of probing the SZ effect and X-ray emission down to a low halo mass regime (such as in a group-like environment, protoclusters, and massive galaxies) and CGM science, allowed by AtLAST, will open a window to connect the cosmic large-scale structure and smaller-scale galactic environment, by informing us of the impact of AGN on a larger scale. In a well-mixed CGM and IGM, the WHIM could spatially coincide with the cold CGM or impact the properties of cold CGM. In this regard, probing spatial scales and metallicities of each component will be key to understanding the baryon mixing and the spatial extent of physical mechanisms in the play. This will be a fundamental step forward in our understanding of galaxy evolution probing a wide range of spatial scales and across cosmic history. We refer readers to our companion paper on the perspective of SZ (and X-ray synergy) for further details (Di Mascolo et al. 2024, in prep.). 

\subsubsection*{The Warm CGM: KCWI/Keck, MUSE/VLT, BlueMUSE/VLT, ELT and GMT}
~~The warm and ionised CGM, extended over several tens or hundreds of kpc, is now frequently detected around bright quasars at all redshifts probed so far thanks to sensitive IFUs with large FoV such as MUSE/VLT \citep{Bacon2010} and KCWI/Keck \citep{Morrissey2018} e.g., \cite{Borisova2016, Cai+19, Fossati2021, Arrigoni-Battaia2019, Johnson2022, Farina2019}. 
The high Ly$\alpha$ surface brightness and the low HeII1640 over Ly$\alpha$ (or H$\alpha$) ratio of these nebulae suggest the presence of a multiphase CGM with high densities and/or broad density distributions, whose origin and properties are still poorly constrained \citep{Cantalupo2019}. In addition, although fainter and with smaller extension, UV and optical emission are also detected around intermediate and high-redshift galaxies \citep{Wisotzki2016, Leclercq2022, Dutta2023}. 
By combining the optical/UV emission lines from the warm CGM with the far-infrared extended emission that will be obtained with AtLAST, we will be able for the first time to directly constrain the density distribution of the warm and cold CGM component using newly developed photo-ionisation models. Moreover, the IFU data and AtLAST will provide a direct image of the morphology and spatial distribution of the multiphase CGM, together with kinematic information. 
New facilities, such as BlueMUSE on VLT \citep{Richard2019} and the 30/40-m class ground-based large opt/NIR telescopes like the Extremely Large Telescope (ELT)\footnote{\url{https://elt.eso.org/}}, which will be available at the same time as the proposed AtLAST, will extend the wavelength range for which comparisons can be made and will provide higher spatial resolution and sensitivity. In particular, the installation of BlueMUSE/VLT will allow us to map lower redshift structures in Ly$\alpha$ and provide a larger FoV than currently available with MUSE/VLT.    

\subsubsection*{Cold CGM in HI and magnetic fields: MeerKAT and SKAO}
The atomic hydrogen (\hi) in the CGM can be investigated by MeerKAT and SKAO with better sensitivity. The \hi\, probe will be mainly driven by nearby, extragalactic communities where the direct detection of \hi\, emission and mapping out large areas of the sky is possible with exquisite sensitivity. The combined studies of \hi\, and molecular gas, the latter probed by AtLAST, will offer insights into the fate of neutral/molecular gas from the large-scale structure (being part of the IGM) onto the galaxy for a complete view of the entire cold CGM flows. Without lensing or statistical analyses, like stacking, studies of atomic hydrogen in the distant universe will be complemented by AtLAST by mapping \cii emissions that may be instead used as a probe of \hi in a distant universe.
The magnetic fields probed by radio continuum emissions can further help to understand the origin of the observed gas dynamics in the CGM (whether this is driven by turbulence made by the magnetic field in the CGM/IGM, e.g., \citealt{Hu2024}).

\subsubsection*{Cold CGM by low CO transitions: Next-Generation VLA (ngVLA)}
Millimetre studies with the next-generation VLA (1.2$-$116\,GHz) will trace the low-$J$ transitions of CO, which are a critical complement to the high-J CO and \ci/\cii lines accessible to AtLAST. By tracing the spectral line energy distribution (SLED), one can derive accurate masses, excitation conditions, and carbon abundances in the molecular CGM at any redshift. At $z$\,$>$\,4, the low-$J$ transitions are significantly dimmed due to the increasing temperature of the Comic Microwave Background (CMB), hence the emission from the CGM may be suppressed \citep[e.g.,][]{dacunha+13,zhang+16}. Operating at higher frequencies, AtLAST will play a key role in measuring gas masses from high-J CO (using a SLED) or using \ci/\cii at the highest redshifts. Because the ngVLA operates at lower mm frequencies than AtLAST, widespread emission in the CGM is less likely resolved out and thus detectable with the compact interferometric array configuration of the ngVLA’s central core \citep{Emonts2018ngvla}. The current VLA (in the North) and ALMA (in the South) have observed the same targets from dec\,=\,-26$^{\circ}$ (Spiderweb) to +40$^{\circ}$ (4C\,41.17), therefore half the sky will accessible to the powerful combination of ngVLA and AtLAST for investigating the physical properties of the molecular CGM.

\section{Summary }\label{sec:summary}
We addressed the need for a new 50-m large FoV sub-mm observatory in the Atacama desert in Chile, called AtLAST, by focussing on the scientific case of probing the (so far, hidden) cold phase of the CGM.
Current theoretical and experimental constraints on the properties of galaxies imply that a significant amount of gas mass is placed beyond the scales of the ISM and within the virial radii of galaxies, at all cosmic epochs. Such a component, which we define as CGM, is the same reservoir that feeds galaxies through accretion and is at the same time altered by the internal feedback mechanisms at work in galaxies. From the current constraints of observations (Section~\ref{sec:obs_constr}) and simulations (Section~\ref{sec:simulations}), an emerging picture is that a fraction of the CGM would be in the form of cold gas ($T\leq10^4$ K) which, because of its low surface brightness and large angular extent, it is consistently missed by currently available sub-mm facilities. Further, the multi-phase nature of the CGM requires targeting multiple emission lines to pin down its physical properties. 

~~\textbf{All of these call for a sensitive (a large aperture) sub-mm single dish telescope with the capability of a large field of view ($\sim$ a degree) and moderately high spectral resolution.} The synergy with other current/upcoming facilities at other wavelengths is also highly anticipated to complete the full picture of the multi-phase nature.
Studying the cold phase CGM with the capabilities allowed by AtLAST will be the next groundbreaking science in understanding galaxy evolution. 
\subsection*{Data and Software availability}
The calculations used to derive integration times for this paper were done using the AtLAST sensitivity calculator, a deliverable of the Horizon 2020 research project ``Towards AtLAST'', and available from \href{https://github.com/ukatc/AtLAST_sensitivity_calculator}{this link}.
The CASA simulator is available as part of the Common Astronomy Software Applications (CASA), an open-source software available through Github under a General Public License \citep{CASA}.

\section*{Competing interests}
No competing interests were disclosed.

\section*{Grant information}
This project has received funding from the European Union’s Horizon 2020 research and innovation program under grant agreement No 951815 (AtLAST).
M. Lee acknowledges support from the European Union’s Horizon Europe research and innovation programme under the Marie Sk\l odowska-Curie grant agreement No 101107795. 
S. Shen acknowledges support from the European High-Performance Computing Joint Undertaking (EuroHPC JU) and the Research Council of Norway through the funding of the SPACE Centre of Excellence (grant agreement No 101093441). 
S. Cantalupo gratefully acknowledges support from the European Research Council (ERC) under the European Union’s Horizon 2020 Research and Innovation programme grant agreement No 864361.
L. Di Mascolo\ is supported by the ERC-StG ``ClustersXCosmo'' grant agreement 716762. L.D.M.\ further acknowledges the financial contribution from the agreement ASI-INAF n.2017-14-H.0. This work has been supported by the French government, through the UCA\textsuperscript{J.E.D.I.} Investments in the Future project managed by the National Research Agency (ANR) with the reference number ANR-15-IDEX-01. 
A. Pensabene acknowledges the support from Fondazione Cariplo grant no. 2020-0902. 
M. Rybak is supported by the NWO Veni project "\textit{Under the lens}" (VI.Veni.202.225).
S. Wedemeyer acknowledges support by the Research Council of Norway through the EMISSA project (project number 286853) and the Centres of Excellence scheme, project number 262622 (``Rosseland Centre for Solar Physics'').


\section*{Acknowledgements}


%

\begingroup
\small
\bibliographystyle{apj_mod}
\setlength{\parskip}{0pt}
\setlength{\bibsep}{0pt}
\bibliography{atlast}
\endgroup

\end{document}